\documentclass[pra,twocolumn,aps,showpacs,superscriptaddress,longbibliography]{revtex4-1}
\usepackage{graphicx}
\usepackage{dcolumn}
\usepackage{bm}
\usepackage{epstopdf}
\usepackage{float}
\usepackage{amsmath}
\usepackage{xcolor}
\usepackage{comment}
\usepackage{hyperref}
\usepackage[normalem]{ulem}
\graphicspath{{./images/}}
\usepackage{pgfplots}

\usepackage{pgfplots}\pgfplotsset{compat=1.8}
\usepackage{tikz}
\usetikzlibrary{calc}
\usepackage{xcolor}
\definecolor{lightblue}{rgb}{0.2,0.2, 0.9}

\begin{document}
\title{Perspective: Interactions mediated by atoms, photons, electrons, and excitons}
\author{Rosario Paredes}
\affiliation{ Instituto de F\'isica, Universidad Nacional Aut\'onoma de M\'exico, Apartado Postal 20-364, Ciudad de M\'exico C.P. 01000, Mexico}
\author{Georg Bruun}
\affiliation{Center for Complex Quantum Systems, Department of Physics and Astronomy, Aarhus University, Ny Munkegade, DK-8000 Aarhus C, Denmark}
\author{Arturo Camacho-Guardian}
\email{acamacho@fisica.unam.mx}
\affiliation{ Instituto de F\'isica, Universidad Nacional Aut\'onoma de M\'exico, Apartado Postal 20-364, Ciudad de M\'exico C.P. 01000, Mexico}

%\date{\today}
\begin{abstract} 
Interactions between quasiparticles mediated by a surrounding environment  are ubiquitous  and lead to a range of important 
effects from collective modes of low temperature quantum gases, superconductivity, to the interaction between elementary 
particles at high energies. This perspective article is motivated by 
experimental progress in the fields of quantum degenerate atomic gases, cavity QED, and 
two-dimensional (2D) semi-conductors, which  enable a systematic exploration of mediated interactions in new settings and 
regimes. We first describe how to microscopically 
calculate the quasiparticle interaction using perturbation theory, diagrammatics, and the path integral, highlighting the 
key role played by the quantum statistics of the quasiparticles. Recent theoretical and experimental insights into 
quasiparticle and mediated interactions in general obtained from atomic gases are then discussed, after which we  
 focus on hybrid light-atom systems where a  remarkable 
long range photon mediated interaction can be realised. Next, we describe new and puzzling 
results regarding the interaction between quasiparticles in 2D semiconductors. 
We then discuss how mediated interactions open up ways to realise new quantum phases in atomic and hybrid atom-photon systems as well as  2D semiconductors, and the perspective ends by posing some open questions and outlook. 
\end{abstract}
\maketitle
%%%%%%%%%%%%%%%%%%%%%%%%%%%%%%%%%%%%%%%%%%%%%%%%%%%%%%%%%%%%
%%%%%%%%%%%%%%%%%%%%%%%%%%%%%%%%%%%%%%%%%%%%%%%%%%%%%%%%%%%%
%%%%%%%%%%%%%%%%%%%%%%%%%%%%%%%%%%%%%%%%%%%%%%%%%%%%%%%%%%%%
%%%%%%%%%%%%%%%%%%%%%%%%%%%%%%%%%%%%%%%%%%%%%%%%%%%%%%%%%%%%

%%%%%%%%%%%%%%%%%%%%%%%%%%%%%%%%%%%%%%%%%%%%%%%%%%%%%%%%%%%%
\section{Introduction}
\begin{figure}[h]
\includegraphics[width=.997\columnwidth]{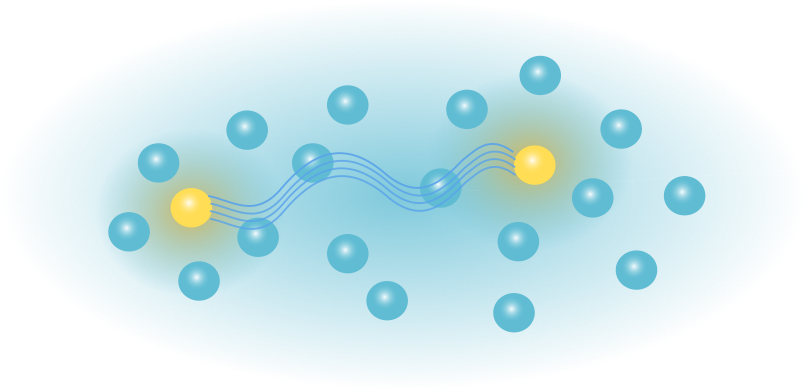}
\caption{Cartoon of mediated interactions. Two impurities (yellow balls) perturb their surroundings, which therefore mediates  an effective interaction between them.} 
\label{Fig0}
\end{figure}

An intrinsic property of quasiparticles is that they interact with each other, because  
the effects of one quasiparticle on its surrounding are felt by another as  illustrated in Fig.~\ref{Fig0}. Such interactions 
 lead to a range of remarkable and important effects including superconductivity caused by an 
attractive phonon mediated interaction between electrons that overwhelms their Coulomb repulsion~\cite{Bardeen1957,Bardeen1957b,Schrieffer1983}, and giant magnetoresistance~\cite{Grunberg1986,Parkin1990}. Another 
 prominent example of mediated interactions is that between $^3$He atoms in liquid $^4$He~\cite{Bardeen1967}. 
Also, it has been conjectured that  interactions between electrons mediated by spin fluctuations are the mechanism behind high-temperature superconductivity \cite{Bednorz1986tc,RKresin2009}, and  at the fundamental level all interactions between elementary particles are mediated 
by gauge bosons \cite{weinberg1995quantum,Gerard1980, LEIKE1999143}.

This perspective is motivated by advances in different experimental platforms 
providing new possibilities to explore mediated interactions in quantum many-body systems. The first platform consists of atomic gases cooled down to very low temperatures such that their properties are governed by 
quantum mechanics. This  realizes highly flexible many-body systems where fundamental properties such as the interaction strenght, spatial 
dimension, and quantum statistics of the particles can be experimentally controlled~\cite{bloch_many-body_2008,Giorgini2008,gross2017quantum}. By inserting such atomic gases in optical cavities, one can moreover realise 
entirely new hybrid light-matter systems with several exotic properties~\cite{mivehvar2021cavity}.
Likewise, atomically thin transition metal dichacogenides (TMDs) like MoSe$_2$, MoS$_2$, WSe$_2$, and WS$_2$
 open up new possibilities to explore truly 2D many-body physics in  a solid state setting.
 They are  direct band gap semiconductors with a rich set of spin-valley degrees of freedom, which
 has led to  the realization of several exotic phases of matter~\cite{regan2022emerging,kennes2021moire}, such as as Mott insulators~\cite{shimazaki2021optical}, Wigner crystals~\cite{regan2020mott,smolenski2021signatures}, topological  superconductivity~\cite{kezilebieke2020topological}, and the integer and fractional quantum  Hall effects~\cite{li2021quantum,xu2023observation}.

In this perspective, we will discuss how these versatile experimental  platforms, which seem quite different but actually share many similarities,
 offer new ways to explore mediated interactions systematically and in regimes never realised before. This has led to new 
 theoretical and experimental understandings as well as opened up new frontiers. The perspective article starts by setting up the 
 basic framework for describing interactions between quasiparticles, and we  then 
 explain  different theoretical approaches for calculating them, from  perturbation 
 theory over diagrammatic theory to the path integral.  This  highlights the key role played by the quantum statistics of the 
 quasiparticles in determining the sign of the interaction, 
 and also the relation between  quasiparticle and general  mediated interactions. 
 We then discus  theoretical and experimental insights regarding 
 interactions mediated in atomic Fermi gases and Bose-Einstein condensates (BECs). The striking realisation of essentially infinite 
 range interactions between atoms mediated by photons in an optical  cavity are then described using both a diagrammatic and an equations of motion approach, after which we turn our attention 
 to the experimental and theoretical advances regarding mediated interactions in TMDs. We end this perspective  
 by summarising  results and discussing some open questions and frontiers.

This article is organized as follows. The fundamental theoretical framework for describing quasiparticles and their interactions is 
described in Sec.\ref{Quasiparticles}. We then explain three different ways to calculate quasiparticle interactions: 
perturbation theory rigorously valid for weak coupling in Sec.\ref{perturbation}, a general diagrammatic 
approach capable of going beying weak coupling in Sec.\ref{Green}, and finally in Sec.\ref{pathI} a path integral approach where mediated 
interactions emerge quite naturally.
The concepts of the quasiparticle and mediated interactions are disentangled in Sec.~\ref{LandauvsInduced}, and 
in Sec.~\ref{fermimediated} we discuss  the recent experimental and theoretical progress regarding 
fermion-mediated interactions in atomic quantum gases. The case of boson-mediated interactions in atomic gases 
is discussed in Sec.~\ref{quantumBosegases}, whereas Sec.~\ref{QOL} is devoted to the long range 
interactions between atoms mediated by cavity photons. In Sec.~\ref{vanderWaals}, we discuss  mediated interactions 
between excitons in TMDs, and the  emergence of collective phases due to mediated interactions are  outlined 
in Sec.~\ref{manybodybose}. We finally end with conclusions and outlook in Sec.~\ref{Conclusions}.

\section{Interactions between quasiparticles}\label{Quasiparticles}
In this section, we will set up the general framework for analysing the interactions between quasiparticles as first put forward by Landau, and  which is a pinnacle in theoretical physics allowing a relatively 
simple yet powerful description of otherwise complicated interaction quantum many-body systems. 
The premise of Landau's theory is that the energy  of a many-body system can be expanded as a power series in the occupation numbers of its fundamental excitations~\cite{baym2008landau}, 
which behave as { either bosonic or fermionic } particle-like entities, so-called quasiparticles, with renormalised properties, finite life-times. Keeping  terms up to second order in the occupation numbers, the energy of the system is written as 
\begin{equation}
\label{landauq}
E=E_0+\sum_{\mathbf p,\sigma}\varepsilon^0_{\mathbf p\sigma}n_{\mathbf p\sigma}+
\frac{1}{2}\sum_{\mathbf p,\mathbf  p',\sigma,\sigma'}f_{\mathbf p\sigma,\mathbf p'\sigma'}n_{\mathbf p\sigma}n_{\mathbf p'\sigma'}~,
\end{equation}
where $E_0$ is the energy of the system in the absence of any quasiparticle excitations, 
$\varepsilon^0_{\mathbf p\sigma}$ is the energy of a single quasiparticle with momentum $\mathbf p$, 
$n_{\mathbf p\sigma}$ is its occupation, and $\sigma$ denotes a possible internal 
quantum state such as spin. We have here and in the rest of the paper assumed a unit system volume. 
The first derivative of Eq.~\eqref{landauq} gives the energy of a quasiparticle 
\begin{equation}
\label{QPenergy}
\varepsilon_{\mathbf p\sigma}=\frac{\delta E}{\delta n_{\mathbf p\sigma}}=\varepsilon_{\mathbf p\sigma}^0+
\sum_{\mathbf  p'\sigma'}f_{\mathbf p\sigma,\mathbf p'\sigma'}n_{\mathbf p'\sigma'}~,
\end{equation}
where the second term is its energy shift due to the interactions with other quasiparticles. The second derivative gives 
the interaction between two quasiparticles in quantum states  $\mathbf p\sigma$ and $\mathbf p'\sigma'$, 
\begin{equation}
\label{landauint}
f_{\mathbf p\sigma,\mathbf p'\sigma'}=
\frac{\delta^2 E}{\delta n_{\mathbf p\sigma}\delta n_{\mathbf p'\sigma'}}.
\end{equation}
Importantly, this interaction is an inherent property of quasiparticles in a medium since one quasiparticle affects its surroundings, which is felt by another quasiparticle. Put another way, two quasiparticles interact by exchanging modulations in their surrounding medium. For a thorough discussion of Landau's quasiparticle theory, we refer 
to Refs.~\cite{baym2008landau,Fetter1971,landau1980}.

\section{Perturbation theory}\label{perturbation}
We present in this section two cases where the quasiparticle interaction given by Eq.~\eqref{landauint} can be calculated rigorously using perturbation theory. 
\subsection{Fermion mediated interaction}\label{Fermipert}
Consider  $N_a$ mobile "impurity" particles of mass $m_a$ immersed in an ideal spin-polarized (single component)  gas of fermionic particles with mass $m_b$
and density $n_F$. The Hamiltonian is $\hat H=\hat H_0+\hat H_\text{int}$ with 
\begin{align}\label{FermiH0}
\hat H_0=&\sum_{\mathbf p}\epsilon_{\mathbf p a} \hat a^\dagger_\mathbf p\hat a_\mathbf p+\sum_{\mathbf p }\epsilon_{\mathbf p b} \hat b^\dagger_{\mathbf p }\hat b_{\mathbf p }~,\\ 
\label{FermiHint}
\hat H_\text{int}=&
g\sum_{\mathbf p,\mathbf p',\mathbf q } \hat a^\dagger_{\mathbf p+\mathbf q}\hat b^\dagger_{\mathbf p'-\mathbf q } \hat b_{\mathbf p' }\hat a_\mathbf p.
\end{align}
Here,  $\hat a^\dagger_\mathbf p$ creates an impurity particle with momentum $\mathbf p$ and kinetic energy 
$\epsilon_{\mathbf p a}=p^2/2m_a$, $\hat b^\dagger_\mathbf p$ creates a fermion with momentum $\mathbf p$ and kinetic energy 
$\epsilon_{\mathbf p b}=p^2/2m_b$, and $g$ is the interaction between the fermions and the impurity particles, which we take to be 
momentum independent (infinitely short range), whereas the fermions do not interact amongst themselves. While such a  system 
may seem oversimplified at first, it has in fact been created experimentally by several groups using cold atomic 
gases and it features several interesting physical effects~\cite{Schirotzek2009,Nascimbene2009,kohstall2012metastability,koschorreck2012attractive,Massignan2014,cetina2016ultrafast,Scazza2017,schmidt2018universal,Fritsche2021}. In particular, the interaction between 
the impurity particles and the  Fermi gas leads to the formation of 
quasiparticles denoted as Fermi polarons, which consist of a impurity particle surrounded by a cloud of fermions.

{Before we delve into the calculations, we note that while Eq.~\ref{FermiHint} relates to the interaction between a bare impurity and a fermion from the Fermi sea, Landau's quasiparticle relates to an effective interaction between quasiparticles, which we now calculate. We begin with the} interaction between two Fermi polarons assuming the interaction $g$ is weak so that perturbation 
theory applies. An eigenstate for no interactions ($g=0$) is  
$|\psi_0\rangle=\prod_{\mathbf p}\hat a^\dagger_{\mathbf p}\prod_{|\mathbf k|<k_F}\hat b^\dagger_{\mathbf k}|0\rangle$ where 
$|0\rangle$ is the vacuum state and $k_F$ is the Fermi momentum. The energy shift to first order is 
$E_1=\langle\psi_0|\hat H_\text{int}|\psi_0\rangle=N_agn_F$, which is simply the mean-field energy 
coming from the density $n_F$  of the fermions. 
The second order energy shift is $E_2=\sum_{n\neq 0}\left|\langle \psi_0|\hat H_\text{int}|\psi_n\rangle\right|^2/(E_0-E_n)$.
The relevant unperturbed eigenstates are 
$|\psi_n\rangle=[n_{\mathbf pa}(n_{\mathbf p-\mathbf qa}+1)]^{-1/2}
\hat a^\dagger_{\mathbf p-\mathbf q}\hat a_{\mathbf p}\hat b^\dagger_{\mathbf k+\mathbf q}\hat b_{\mathbf k}|\psi_0\rangle$ where  $n_{\mathbf pa}$ is the occupation function of the impurities. This straightforwardly 
gives 
\begin{equation}
\label{secondorderFermions}
E_2=g^2\sum_{\mathbf p,\mathbf k,\mathbf q}\frac{n_{\mathbf kb}(1-n_{\mathbf k+\mathbf q b})n_{\mathbf pa}(1\pm n_{\mathbf p-\mathbf qa})}{\epsilon_{\mathbf kb}+\epsilon_{\mathbf pa}-\epsilon_{\mathbf k+\mathbf qb}-\epsilon_{\mathbf p-\mathbf q a}},
\end{equation}
where $n_{\mathbf kb}=1/\{\exp[(\epsilon_{\mathbf kb}-\mu_b)/T]+1\}^{-1}$ is the occupation function for the fermions with 
chemical potential $\mu_b$. The upper/lower sign in the second bracket in Eq.~\eqref{secondorderFermions} is for 
bosonic/fermionic impurities and originates from  Bose stimulation/Fermi blocking 
of the scattering processes. {Mathematically, it arises from the (anti)-commutation rules for (fermions) and bosons.}

From Eqs.~\eqref{landauint} and \eqref{secondorderFermions}, we then get~\cite{Yu2012}
\begin{equation}
\label{RKKY}
f_{\mathbf p,\mathbf p'}=\pm g^2\chi(\mathbf p-\mathbf p',\epsilon_{\mathbf pa}-\epsilon_{\mathbf p'a}),
\end{equation}
for the quasiparticle interaction between two Fermi polarons with momenta $\mathbf p$ and $\mathbf p'$ to second order in $g$.
Here 
\begin{equation}
\label{Lindhard}
\chi(\mathbf p,\omega)
=\sum_{\mathbf k}\frac{n_{\mathbf kb}-n_{\mathbf k+\mathbf p b}}{\omega+\epsilon_{\mathbf kb}-\epsilon_{\mathbf k+\mathbf pb}},   
\end{equation}
is the density-density response function of an ideal Fermi gas (the Lindhard 
function) describing how compressible it is~\cite{Giuliani2005} and we remind the reader that the upper/lower sign is for bosonic/fermionic impurities. 
In the limit of small momentum exchange and zero temperature, the quasiparticle interaction becomes 
\begin{equation}\label{RKKYzeromomentum}
\lim_{|\mathbf p|\rightarrow |\mathbf p'|}f_{\mathbf p,\mathbf p'}=\mp g^2\mathcal N(\epsilon_F)=\mp\frac{(\Delta N)^2}{\mathcal N(\epsilon_F)},
\end{equation}
where  $\mathcal N(\epsilon_F)$ is the density of single-particle states at the Fermi energy $\epsilon_F$~\cite{Yu2012,Yu2010,Mora2010,Giraud2012}. These expressions are general for any dimensionality of the system. In particular, for 3D, $\mathcal N(\epsilon_F)=m_ck_F/2\pi^2$. In the second equality of Eq.~\ref{RKKYzeromomentum}, we have expressed the interaction as $g=(\partial \mu_2/\partial n_1)_{n_2}$, which can be related to the number $\Delta N$ of fermions attracted/repelled by the impurity~\cite{Yu2012,Yu2010}. 
Equation \eqref{RKKYzeromomentum} holds as long as one can perform perturbation theory in the interaction between Fermi polaron and its surroundings.  

This derivation explicitly demonstrates the crucial role played by the quantum statistics of the quasiparticles: 
 for \emph{fermionic} impurities the quasiparticle 
interaction is \emph{repulsive} whereas it is \emph{attractive} for \emph{bosonic} impurities. The origin of this sign difference is the Fermi blocking or Bose stimulation of impurity scattering on the fermions due to the 
presence of other impurities.

This calculation also  illustrates how it 
the interaction between two quasiparticles is mediated by the exchange of particle-hole excitations in the Fermi sea.

\subsection{Boson mediated interaction}\label{Bosepert}
As a second example where rigorous results are possible, we  consider $N_a$ mobile impurities in a weakly interacting Bose gas forming a Bose-Einstein condensate (BEC) at zero temperature. Using Bogoliubov theory to describe the BEC gives the Hamiltonian 
\begin{gather}
\hat H=\sum_{\mathbf p}(\epsilon_{\mathbf p a}+n_0g) \hat a^\dagger_\mathbf p\hat a_\mathbf p+\sum_{\mathbf p} E_{\mathbf p}\hat\gamma^\dagger_{\mathbf p}\hat\gamma_{\mathbf p}+
\nonumber\\ 
+g\sum_{\mathbf p,\mathbf q}\hat a^\dagger_{\mathbf p+\mathbf q}\hat a_\mathbf p\left[\sqrt{\frac{n_0\epsilon_{\mathbf qb}}{E_{\mathbf q}}}(\hat\gamma^\dagger_{-\mathbf q}+\hat\gamma_{\mathbf q})+\sum_{\mathbf k}\hat b^\dagger_{\mathbf k-\mathbf q}\hat b_{\mathbf k}\right],
\label{Hboson}
\end{gather}
where $\hat \gamma_\mathbf p^\dagger=u_{\mathbf p} {\hat b}_{\mathbf b}^\dagger+v_{\mathbf p} {\hat b}_{-\mathbf p}$ 
creates a Bogoliubov excitation in the BEC with energy $E_{\mathbf p}=\sqrt{\epsilon_{\mathbf pb}(\epsilon_{\mathbf pb}+2n_0g_{b})}$,
and $u^2_{\mathbf p}=1+v^2_{\mathbf p}=[(\epsilon_{\mathbf pb}+n_0g_{b})/E_{\mathbf p}+1]/2$ are the usual 
coherence factors, and $n_0$ the condensate fraction. Here ${\hat b}_{\mathbf p}^\dagger$ creates a boson with momentum $\mathbf p$ and $g_{b}$ is the 
 strength of the short range repulsion between the bosons.
 We have in Eq.~\eqref{Hboson} omitted any constant terms giving the  energy 
of the BEC ground state  and all momenta for the bosons are different from zero. 
The Bogoliubov energy is linear at small momentum, $E_{\mathbf p}\approx c_sp$ with $c_s=\sqrt{n_0g_{b}/m_b}$ the BEC speed of sound,
which resembles the acoustic phonon dispersion in solids. The last term describes interactions between the impurities and the BEC 
leading to the formation of  quasiparticles  coined {\it Bose polarons.}

In the perturbative regime where  both $g$ and $g_b$ are small, one can neglect the last quartic
term in Eq.~\eqref{Hboson} since its contribution to the quasiparticle interaction is suppressed by the small depletion of the BEC. For instance, in 3D, this suppression is proportional to $(n_0a_b^{1/3})^{1/2}$~\cite{camacho2018landau}, where
$a_b$ is the boson-boson scattering length with $g_{b}=4\pi a_b/m_b$. 
The  interaction can now be calculated using second order perturbation 
theory following the same steps as in Sec~\ref{Fermipert}. %, where the relevant excited states are now  
 
 This gives the second order energy shift~\cite{Yu2012}
\begin{align}
\label{secondorderBosons}
E_2=%-\frac{g^2}{V}
g^2\sum_{\mathbf p,\mathbf q}\frac{n_0\epsilon_{\mathbf qb}}{E_{\mathbf q}}&\left[\frac{
(1+n_{\mathbf qb})n_{\mathbf pa}(1\pm n_{\mathbf p-\mathbf qa})}
{\epsilon_{\mathbf pa}-\epsilon_{\mathbf p-\mathbf qa}-E_{\mathbf q}}\right.\nonumber\\ 
&\left.+\frac{n_{-\mathbf qb}n_{\mathbf pa}(1\pm n_{\mathbf p-\mathbf qa})}
{\epsilon_{\mathbf pa}+E_{-\mathbf q}-\epsilon_{\mathbf p-\mathbf qa}}\right],
\end{align}
where $n_{\mathbf qb}=1/[\exp(E_{\mathbf q}/T)-1]^{-1}$ is the occupation function for the bosons
and the upper/lower sign again is for bosonic/fermionic impurities.

Equations \eqref{landauint} and \ref{secondorderBosons} then again gives Eq.~\eqref{RKKY} for the quasiparticle interaction, but now with
\begin{equation}
\chi(\mathbf p,\omega)=\frac{2n_0\epsilon_{\mathbf pb}}{\omega^2-E_{\mathbf p}^2}
\label{compressibilitybosons}
\end{equation}
the density-density response function of a weakly interacting BEC~\cite{Yu2012}. 

In the limit of small momentum transfer, we obtain

\begin{equation}
\label{MediatedIntBEClowmomentum}
\lim_{|\mathbf p|\rightarrow |\mathbf p'|}f_{\mathbf p,\mathbf p'}=\mp \frac{g^2}{g_{b}}.
\end{equation}
Physically, the $1/g_b$ dependence in Eq.~\eqref{MediatedIntBEClowmomentum} reflects that the interaction increases with the 
compressibility of the BEC. {The second-order contribution in Eq.~\ref{secondorderBosons} arising from the quartic term in Eq.~\ref{Hboson} is suppressed by the gas factor $n_0a^3\ll 1$ ~\cite{Levinsen2017} and gives a small correction to the induced interaction, except for very low momenta  where it gives an unphysical divergence for $\lim_{|\mathbf p|\rightarrow |\mathbf p'|}f_{\mathbf p,\mathbf p'}.$
}

Like 
for the fermionic case discussed in Sec.~\ref{Fermipert}, this derivation explicitly demonstrates the key role of  quantum 
statistics making the quasiparticle interaction attractive/repulsive for bosonic/fermionic quasiparticles.

\section{Field theory}\label{FieldTheorySec}
Having derived rigorous expressions for the interaction between quasiparticles mediated by a 
Fermi or Bose gas in the perturbative regime and 
demonstrated the key role of quantum statistics, we now discuss how this quasiparticle interaction is calculated in a field theoretical approach. 
This provides a framework to at least formally describe the regime of strong interactions and to introduce the induced interaction in general settings 
beyond Landau's quasiparticle interaction.

\subsection{Green's function formalism}
\label{Green}
Green's functions provide a compact  way to analyse interacting many-particle systems~\cite{Fetter1971,abrikosov2012methods}. 
Consider again the $N_a$ impurity particles of mass $m_a$ in a medium of majority particles. The {retarded} Green's function of an impurity with momentum $\mathbf p$ is given by $G_a(\mathbf p,\omega)$ and it is governed by the Dyson's equation $G_a^{-1}(\mathbf p,\omega)=G^{-1}_{a0}(\mathbf p,\omega)-\Sigma_a(\mathbf p,\omega)$ where $G^{-1}_{a0}(\mathbf p,\omega)=\omega-\epsilon_{\mathbf pa}$ is the noninteracting Green's function and $\Sigma_a(\mathbf p,\omega)$ the self-energy of the impurity atoms.  The energy of the 
quasiparticle state of an impurity with momentum $\bf p$ is determined by the self-consistent 
equation~\cite{Fetter1971,abrikosov2012methods}
\begin{equation}
\label{EqQ}
\varepsilon_{\mathbf p}=\epsilon_{\mathbf pa}+\text{Re} \Sigma_a(\mathbf p,\varepsilon_{\mathbf p}).
\end{equation}
Comparing with Eq.~\eqref{QPenergy}, we see in order to extract the quasiparticle interaction, we must calculate how 
the self-energy depends on the impurity concentration. Explicitly, we obtain from Eqs.~\eqref{QPenergy} and \eqref{EqQ}
%Landau's effective interaction can be written as a first functional derivative of the self-energy with respect to the density of $c$ atoms 
\begin{equation}
\label{landausigma}
f(\mathbf p,\mathbf p')=\frac{\delta \varepsilon_{\bf p}}{\delta n_{\bf p'}}=Z_{\mathbf p}\frac{\delta \text{Re}\Sigma_a(\mathbf p,\varepsilon_{\bf p})}{\delta n_{\mathbf p'}},
\end{equation}
where 
\begin{gather}
Z_{\mathbf p}=\left.\left[1-\frac{\partial \text{Re}\Sigma_{a}(\mathbf p,\omega)}{\partial \omega}\right]^{-1}\right|_{\omega=\varepsilon_{\mathbf p}},
\end{gather}
is the quasiparticle residue and we have ignored the spin index $\sigma$ for simplicity. The presence of the residue reflects that it is only the quasiparticle part of the Green's function that contributes to the quasiparticle  
interaction. Note that Eq.~\eqref{landausigma} is generally valid and not restricted to the perturbative regime. We now give two   
examples for calculating the self-energy and the resulting effective interaction from Eq.~\eqref{landausigma}.

\begin{figure}
\includegraphics[width=1\columnwidth]{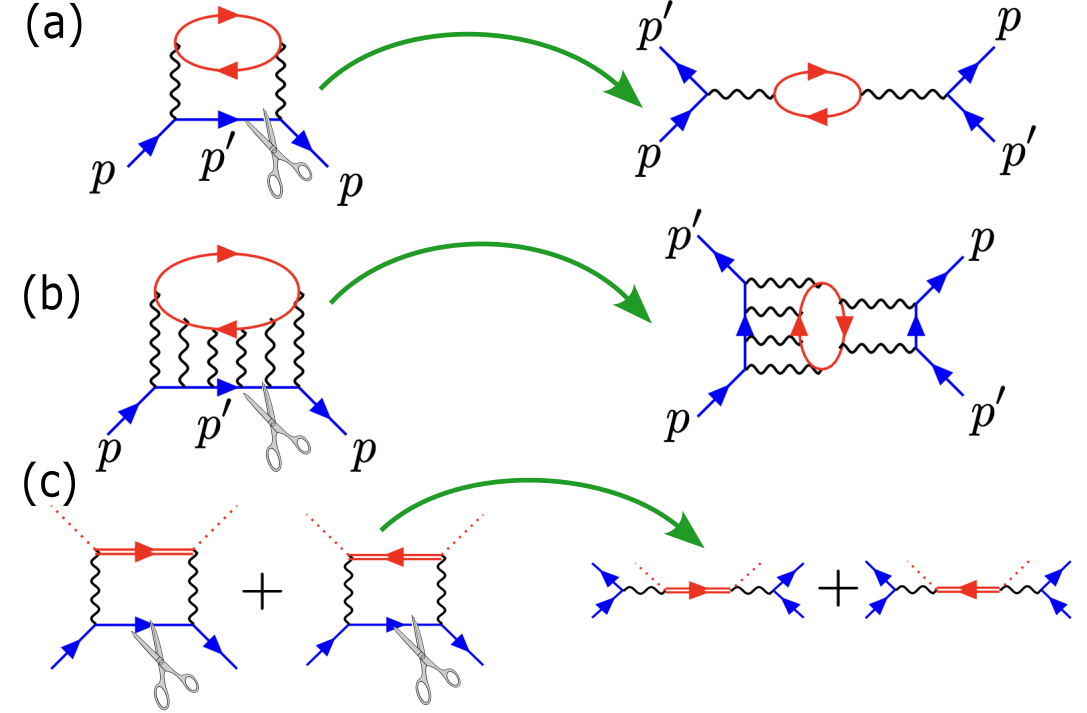}
\caption{(a) Second-order self-energy  of an impurity in a Fermi gas. The derivative
in Eq.~\eqref{landausigma}   corresponds to cutting the intermediate line and gives rise to the 
quasiparticle interaction shown on the right.  The solid red/blue lines correspond to the bath/impurity Green's function, the wavy black line depicts the  impurity-fermion interaction $g$. (b) A typical higher-order process in the ladder approximation for an impurity in a Fermi gas. Cutting an impurity propagator yields the diagram for the quasiparticle  interaction shown on the 
right. (c) Second-order self-energy  for an impurity immersed in a BEC. Cutting the intermediate impurity line leads to a phonon-mediated interaction between the impurities (right). Here the red solid double lines correspond to a Bogoliubov mode propagator. The dashed lines correspond to condensate bosons.
} 
\label{Fig1}
\end{figure}

First, consider again the case described in Sec.~\ref{Fermipert}, i.e.\ 
mobile impurity particles in a single component Fermi gas described 
by the Hamiltonian in Eqs.~\eqref{FermiH0}-\eqref{FermiHint}. The self-energy of the impurities is simply given by the 
mean-field shift $\Sigma_{1a}=g n_F$ to first order in $g$, which is independent of the impurity concentration. The second order self-energy shown in Fig.~\ref{Fig1}(a) is  
\begin{equation}
\Sigma_{2a}({\bf p},i\omega_p)=\pm g^2\int\!\frac{d^3p'}{(2\pi)^3}\chi({\bf p}-{\bf p}',\omega)n_{{\bf p}'a},
\label{SecondOrderSelf}
\end{equation}
where the upper/lower sign is for bosonic/fermionic impurities with the distribution function $n_{{\bf p}a}$ and we have ignored terms independent on the impurity concentration. 
Taking the derivative of $\Sigma_{2a}$ according to Eq.~\eqref{landausigma} then recovers  Eqs.~\eqref{RKKY}-\eqref{Lindhard} for the Landau quasiparticle 
interaction between two quasiparticles mediated by a Fermi gas. As illustrated in Fig.~\ref{Fig1}, taking the derivative corresponds diagrammatically to cutting 
the impurity line inside the self-energy. Here, the blue/red lines correspond to impurity/fermion propagators, whereas the wavy lines are the direct impurity-boson interaction. This calculation also shows that the sign difference between the  effective interaction
for fermonic and bosonic impurities can be understood from the fact that the mediated interaction originates from an 
exchange (Fock) diagram~\cite{Yu2012,Mora2010,Yu2010}.

The power of the Green's function method is that provides a formalism  to analyse the strong coupling regime beyond perturbation theory. 
An example of this is shown in Fig.~\ref{Fig1}(b), where the so-called ladder diagrams are included to infinite order in the self-energy. This approach, which describes two-body correlations between the impurities and the fermions exactly, is known to be remarkably accurate for mobile impurities interacting  with a Fermi 
gas via a short range potential, i.e.\ precisely the problem at hand~\cite{Massignan2014}. Diagrammatically, the corresponding quasiparticle interaction 
is obtained  by cutting one of the internal impurity lines in the self-energy due to the derivative in Eq.~\eqref{landausigma}, which is illustrated in Fig.~\ref{Fig1}(b). As we shall discuss in 
Sec.~\ref{fermimediated}(b), the resulting quasiparticle interaction agrees well with  experimental results for the interaction between Fermi polarons in an atomic gas. 

As a second example, consider again impurities in a weakly interacting BEC as described  in Sec.~\ref{Bosepert}. The first order diagram gives the mean field shift as before and  the second order self-energy is shown diagrammatically in Fig.~\ref{Fig1}(c). The impurity/Bogoliobov mode propagators are depicted by the blue/double red lines respectively. Condensate bosons are illustrated by the red dashed lines. Cutting the impurity line (taking the derivative in Eq.~\eqref{landausigma}), 
gives Eq.~\eqref{SecondOrderSelf} with the compressibility of a BEC given by Eq.~\eqref{compressibilitybosons}. This in turn recovers  
Eq.~\eqref{RKKY} for the effective interaction. Again, this perturbative result can be generalised to strong interactions 
using the ladder approximation as described in Sec.~\ref{quantumBosegases} and Ref.~\cite{camacho2018landau}.

\subsection{Path integral approach}
\label{pathI}
Mediated interactions emerge naturally within the path integral formulation of quantum field theory~\cite{Bijlsma2000,stoof2009ultracold}. 
We illustrate this by considering again the case of the interaction between impurity particles mediated by a weakly interacting BEC as described 
by the Hamiltonian in Eq.~\eqref{Hboson}; a similar calculation for the case of a fermion mediated interaction can be found in e.g.\ Refs.~\cite{Zheng2021,Zheng2024}. The starting point is  
the partition function $Z=\text{Tr} e^{-\beta ({\hat H}-\mu {\hat N})}$ written as a path integral~\cite{NegeleOrland1998}
\begin{equation}
Z=\int \!{\mathcal D}(\gamma^*,\gamma,a^*,a)e^{-S(\gamma^*,\gamma,a^*,a)},
\end{equation}
where the action  is 
\begin{gather}
S=\sum_{p} \left[a^*_p(-i\omega_p+\epsilon_{\mathbf p a}+n_0g) a_p+\gamma^*_{p}(-i\omega_p+E_{\bf p})\gamma_{p}\right]\nonumber\\ 
+g\sum_{p,q} a^*_{p+q}a_p\sqrt{\frac{n_0\epsilon_{\mathbf qb}}{E_{\mathbf q}}}(\gamma^*_{-q}+\gamma_{q}).
\end{gather}
Here, $\gamma_{p}$ are complex numbers in momentum/frequency space $p=({\bf p},i\omega_p)$ with $\omega_p$ a bosonic 
Matsubara frequency describing the Bogoliubov modes in the BEC, and $a_{p}$ are complex/Grassman numbers depending on whether 
the impurities are bosons/fermions. 
We have omitted the quartic term in Eq.~\eqref{Hboson} as its contribution to the induced interaction is suppressed within Bogoliubov theory, for 3D by a factor 
$(n_0a_b^{1/3})^{1/2}$ as discussed in Sec.~\ref{Bosepert}~\cite{camacho2018landau}.

Since there are only quadratic terms for the complex Bose field $\gamma_p$, it is straightforwardly  integrated out using the 
Gaussian identity~\cite{NegeleOrland1998}
\begin{equation}
\int\! {\mathcal D}(\gamma_p^*,\gamma_p)e^{\gamma^*_{p}G_B^{-1}(p)\gamma_{p}-J_p^*\gamma_p-\gamma_p^*J_p}\propto 
e^{-J_p^*G_B(p)J_p},
\end{equation}
where $G_B(p)=1/(i\omega_p-E_{\bf p})$ is the Green's function for the Bogoliubov modes and 
$J_p=g\sqrt{n_0\epsilon_{\mathbf pb}/E_{\mathbf p}}\sum_{q} a^*_{p+q}a_q$.
Using this, we get that the effective action for the impurities is 
\begin{gather}
S_\text{eff}=\sum_{p} a^*_p(-i\omega_p+\epsilon_{\mathbf p a}+n_0g) a_p+\nonumber\\ 
+\frac{g^2}2\sum_{p,q,q'} \frac{2n_0\epsilon_{{\bf p}b}}{(i\omega_p)^2-E_{\bf p}^2}a^*_{q+p}a_qa^*_{q'-p}a_{q'}.
\label{Seff}
\end{gather}
From this we see that, {after analytic continuation $i\omega_p\rightarrow \omega+i0^+$,} the BEC mediates a frequency dependent interaction $g^2\chi(\bf p,\omega)$ (see Eq.~\eqref{compressibilitybosons}) between the impurities  
 with the compressibility of the BEC given by Eq.~\eqref{compressibilitybosons}. Calculating the second order Fock diagram with this interaction 
then gives Eq.~\eqref{SecondOrderSelf} from which  Eq.~\eqref{RKKY} for the  quasiparticle interaction  again is recovered.  {Here again, we ignore the quartic term in the impurity-boson coupling, which is expected to give small corrections.}

\section{Quasiparticle  interaction versus induced interaction}
\label{LandauvsInduced}
The  analysis in Sec.~\ref{FieldTheorySec} shows  that the induced interaction emerges naturally in many-body systems and is not 
restricted to the interaction between quasiparticles. This is  explicit in Eq.~\eqref{Seff}, where an induced interaction
 $V_\text{ind}({\bf p},\omega)=g^2\chi({\bf p},\omega)$ between the 
impurities mediated by the BEC appears already in the effective Lagrangian. 
Indeed, the induced interaction is quite generally  given by 
$V_\text{ind}({\bf p},\omega)=g^2\chi({\bf p},\omega)$ in the perturbative regime, where $\chi({\bf p},\omega)$ is the compressibility of the medium carrying the interaction. This 
expression can be  generalised to strong interactions by considering the appropriate diagrams as exemplified  in Sec.~\ref{Green}. 
The frequency dependence of the induced interaction reflects that excitations in the 
medium carrying the interaction propagate with a finite speed giving rise to retardation effects. Such retardation effects are very important for instance 
in the so-called strong coupling superconductors~\cite{Mahan2000book}. 

The induced interaction and Landau's quasiparticle interaction are connected by
\begin{align}
 f(\mathbf p,\mathbf p')=\pm V_{\text{ind}}(\mathbf p-\mathbf p',\varepsilon_{\mathbf p'}-\varepsilon_{\mathbf p})  
\end{align}
where the upper/lower sign is for bosons/fermions.

It is illuminating to consider the induced interaction in the static limit ($\omega\rightarrow 0)$. 
Fourier transforming Eq.~\eqref{compressibilitybosons} to real space yields in 3D 
\begin{gather}
\label{VYukawa}
V_{\text{med}}(\mathbf r)=-\frac{g^2n_0m_B}{\pi}\frac{e^{-\sqrt{2}r/\xi}}{r}.
\end{gather}
Equation~\eqref{VYukawa} is the well-known Yukawa interaction  
between two static impurities separated by a distance $r$ in a BEC. Note that this interaction is always attractive as opposed to the quasiparticle interaction, where the sign depends on the quantum statistics of the impurities. The reason is that two static impurities 
separated by a distance are distinguishable so that  quantum statistics is irrelevant. Physically, the attraction arises because one impurity 
increases/decreases the density of the BEC in its neighborhood for attractive/repulsive impurity-boson interaction, which attracts another impurity. 
The range of the Yukawa interaction in Eq.~\eqref{VYukawa} is given by the BEC healing length $\xi=1/\sqrt{2m_b n_bg_{b}}$, which 
 increases with decreasing $g_b$
reflecting that a softer (more compressible) BEC is more efficient at mediating an interaction as discussed in 
Sec.~\ref{Bosepert}. 

Likewise, the interaction between two static impurities in a three-dimensional single component Fermi gas at zero temperature is  given by 
\begin{gather}
\label{RKKYrealaspace}
V_{\text{med}}(\mathbf r)=g^2\frac{m_b}{16\pi^3}\frac{2k_Fr\cos(2k_Fr)-\sin(2k_Fr)}{r^4},
\end{gather}
within second order perturbation theory, which is the well-known Ruderman–Kittel–Kasuya–Yosida (RKKY) interaction~\cite{Ruderman1954,Kasuya1956,Yosida1957}.

\section{Quantum Gases: fermion-mediated interactions}\label{fermimediated}
Having set up the general framework to describe and calculate mediated interactions including that between 
quasiparticles using  perturbation theory, as well as many-body theory based on 
Feynman diagrams and the path integral, we now turn to describing the experimental and theoretical advances in recent years regarding this using quantum degenerate atomic gases as quantum simulators. 
A very powerful feature of atomic gases is that the impurity-medium interaction can be tuned using Feshbach resonances, which  allows one to explore mediated interactions systematically and in 
regimes never realised before. 
We first discuss interactions mediated by fermions after which we turn our attention to interactions mediated by a BEC.

\subsection{Mediated interactions in 3D}%{\bf Polaron interactions in 3D.-}
Two experiments based on atomic Bose-Fermi mixtures measured a fermion mediated interaction between bosons in a BEC in the weak coupling regime. 
In the first experiment, the presence of a surrounding quantum degenerate $^6$Li gas 
 was observed to decrease the radius of a $^{133}$Cs BEC 
and also give rise to a collapse instability~\cite{desalvo2019observation}. Due to the small mass of $^6$Li atoms compared to that of $^{133}$Cs atoms, the frequency dependence of 
the mediated interaction was expected to be weak and the results were
consistent with an attractive interaction between the bosons given by the perturbative result Eq.~\eqref{RKKYzeromomentum}. A second experiment
used Ramsey  spectroscopy to measure the frequency shift of a hyperfine transition for bosonic $^{87}$Rb atoms immersed in a  degenerate $^{40}$K
Fermi mass~\cite{Edri2020}. The experimental results were  found to agree with a mean-field shift from a mediated interaction given by  Eq.~\eqref{RKKYzeromomentum}.

A recent experiment systematically probed the quasiparticle interaction between Fermi polarons formed by $^{40}$K or $^{41}$K atoms in a degenerate Fermi gas of $^6$Li atoms~\cite{baroni2024mediated}. 
Using RF spectroscopy, the polaron energy was measured as a function of the quasiparticle particle concentration $n$ and compared to a momentum average of Eq.~\eqref{QPenergy}, i.e.\ 
$\varepsilon=\varepsilon^0+\bar fn$, since experiment had no momentum resolution.  The quasiparticle interaction 
was explored well beyond the perturbative regime using a Feshbach resonance to tune the K-Li interaction.  
 Moreover, by selecting either fermionic $^{40}$K or bosonic $^{41}$K atoms leaving everything else  essentially unchanged, the effects of quantum statistics could be investigated directly.

\begin{figure}
\includegraphics[width=0.99\columnwidth]{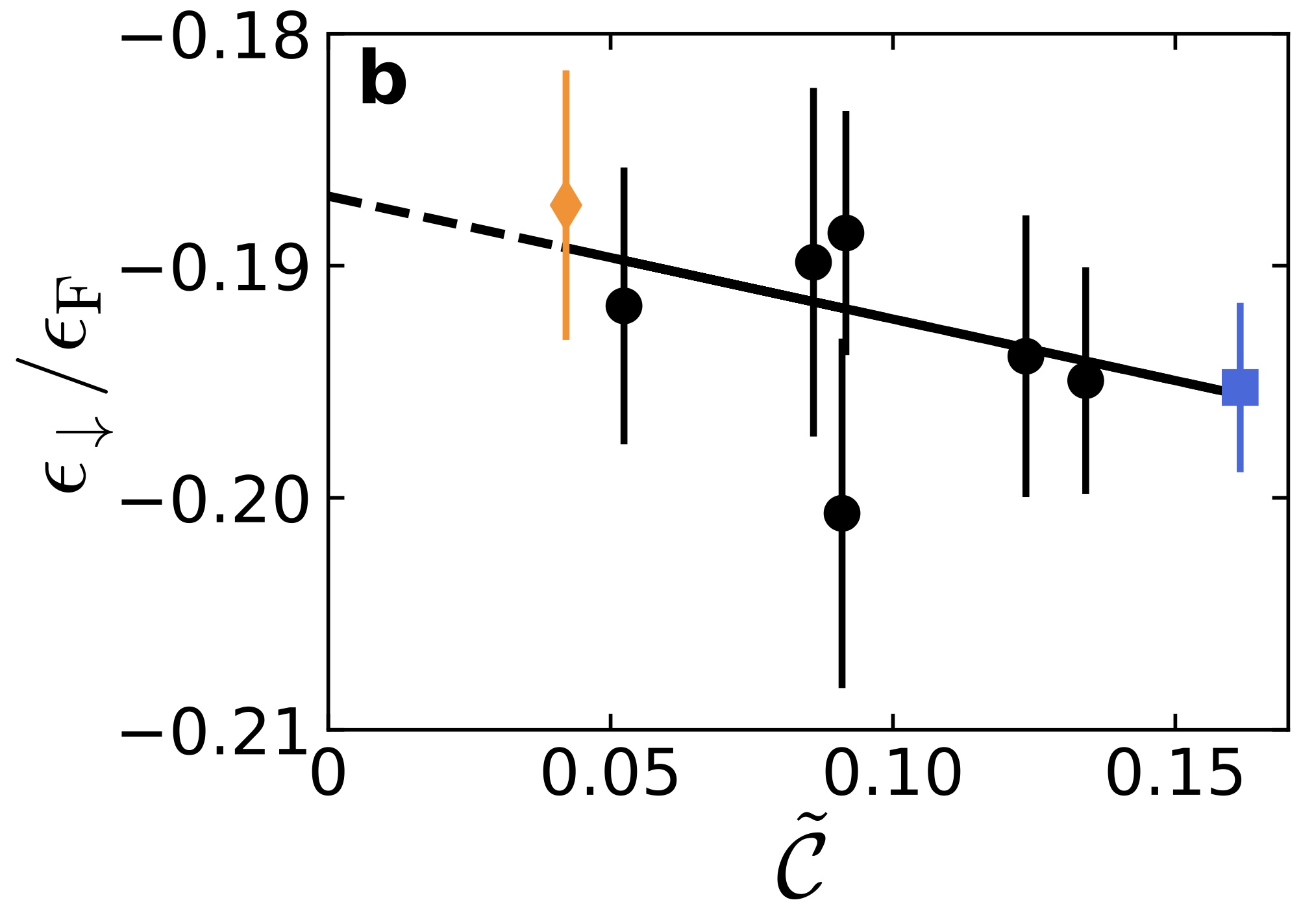}
\includegraphics[width=0.99\columnwidth]{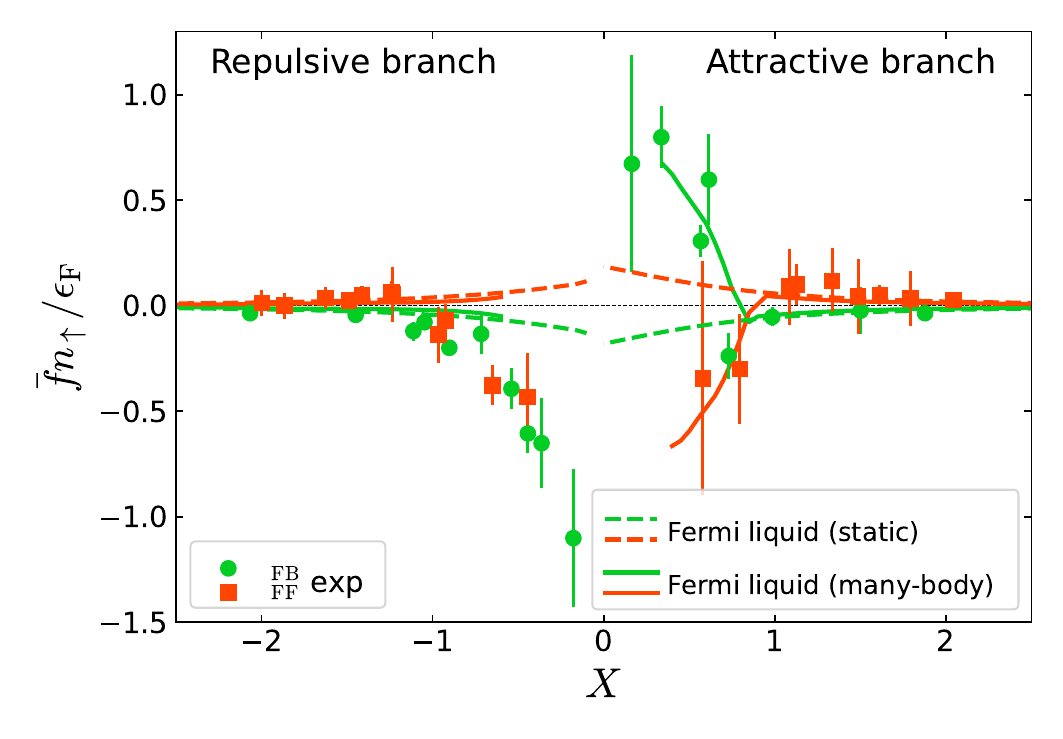}
\caption{Figures from Ref.~\cite{baroni2024mediated}. Upper  panel: The polaron energy as a function of its concentration
for  $^{41}$K atoms with Li-K interaction strength $X=-1/k_Fa=0.98$. Lower panel: The polaron-polaron interaction 
for fermionic $^{40}$K impurity atoms (red squares) and bosonic $^{41}$K impurity atoms (green circles). The dashed lines are the 
perturbative result Eq.~\eqref{RKKYzeromomentum} and the solid lines are a strong coupling theory for the quasiparticle interaction shown diagrammatically in Fig.~\ref{Fig1}(b). 
}
\label{GrimmFig}
\end{figure}
An example of such a measurement of the polaron energy as a function of its concentration is shown in the upper  panel of Fig.~\ref{GrimmFig} for the case of $^{41}$K impurities. The Li-K interaction 
is moderately strong with a scattering length $a$ giving $1/k_Fa=-0.98$, where $k_F$ is the Fermi momentum of the $^6$Li gas. As can be seen, the energy of the polaron decreases with 
increasing concentration corresponding to an attractive average interaction between the polarons, which can be extracted from the slope. The lower panel of 
Fig.~\ref{GrimmFig} shows the interaction $\bar f$ 
 obtained by carefully performing such measurements for different scattering lengths and concentrations, both for $^{40}$K (red points) and $^{41}$K (green points) impurities. 
The first thing to notice is that the quasiparticle interaction is repulsive/attractive for fermionic/bosonic impurities thereby   showing the key role of quantum statistics discussed in Sec.~\ref{perturbation}. Second, 
the strength of the effective interaction agrees well with the second order result given by Eq.~\eqref{RKKYzeromomentum} (dashed lines)
for weak to moderately strong Li-K interaction 
without any fitting parameters. The experiment thus confirms two predictions of Landau's Fermi liquid theory: The strength of the quasiparticle interaction and its sign dependence on the quantum statistics of the  quasiparticles.
The perturbative result for the quasiparticle interaction however breaks down  strong Li-K interactions 
across  the Feshbach resonance where $a$ diverges as can be seen in Fig.~\ref{GrimmFig}.
Here, a theory  for the quasiparticle interaction including retardation effects and Feshbach molecule formation (solid lines)
was able to explain the results for strong and attractive Li-K interactions (negative $a$). It is based on 
Eq.~\eqref{landausigma} together with a ladder approximation for the impurity self-energy generalised to a non-zero impurity concentration  and is shown diagrammatically in Fig.~\ref{Fig1}(b).
The experimental results for strong and repulsive interactions were  not recovered by this theory, which 
may be due to the break-down of Fermi liquid theory or the formation of attractive instead of repulsive polarons, and requires further analysis. 

\begin{figure}
\includegraphics[width=0.99\columnwidth]{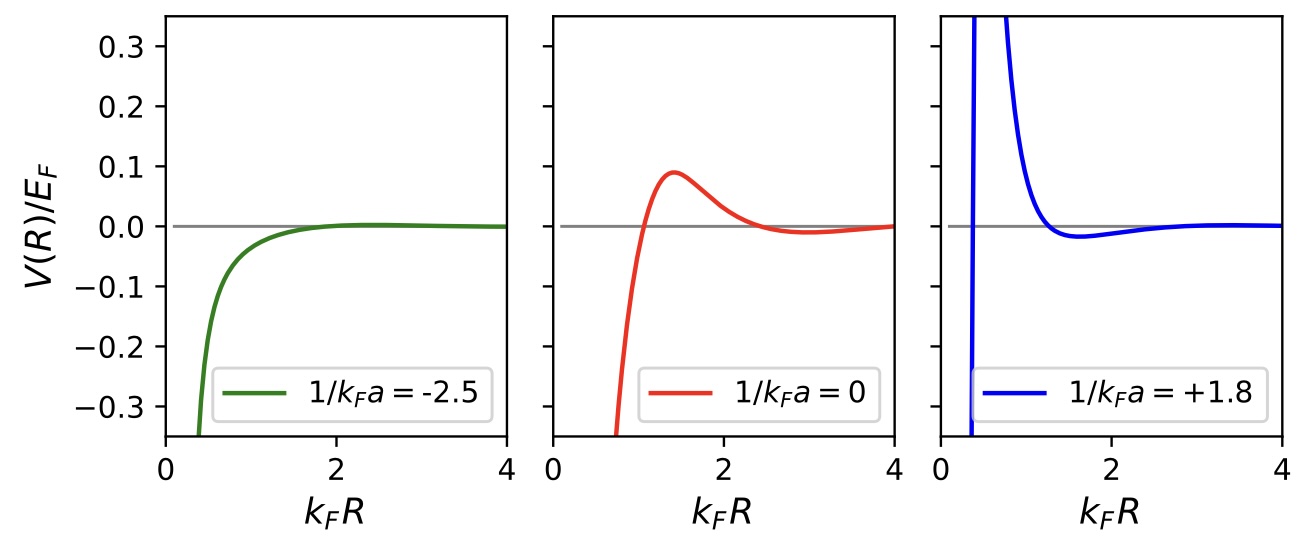}
\caption{Figure from Ref.~\cite{enss2020scattering}. 
Mediated interaction between two static impurities in a Fermi gas for different impurity fermion scattering lengths. 
} 
\label{EnssFig2}
\end{figure}

The interaction between two static (immobile) impurities is simpler to explore than between mobile impurities, since they can be 
treated as scattering potentials without any recoil. This limit 
should  correspond to an infinite impurity mass apart from the fact that  the role of quantum statistics 
is lost since the localised impurities are distinguishable. This case is analogous to the Casimir force between 
 two parallel conducting plates separated by a distance $r$ much smaller than the sides of the mirrors. The vacuum fluctuations of the electromagnetic field depend on the boundary conditions imposed by the plates and therefore, they are different with or without plates, leading to what is referred to as Casimir energy, scaling as $1/r^3$~\cite{Casimir-1948}. 
The interaction between two static impurities mediated by an ideal Fermi gas was studied theoretically in Refs.~\cite{Nishida2009,enss2020scattering}. By solving the scattering problem of   two short range interaction 
potentials exactly, it was shown that interaction can be very different from the RKKY form for strong interactions 
due to the presence of (Efimov) states, where one fermion is bound to the two impurities, see Fig.~\ref{EnssFig2}. 
These bound states lead to resonances and sign changes in the scattering length~\cite{enss2020scattering}. For weak interactions, the perturbative 
RKKY interaction given by Eq.~\eqref{RKKY} was recovered.
%%%%%%%%%%%%%%%%%%%%%%%%%%%%%%%%%%%%%%%%%%%%%%%%%%%%%%%%%%%%%%%%%%%%%%%%%%%%%%%%%%%%%%%%%
%%%%%%%%%%%%%%%%%%%%%%%%%%%%%%%%%%%%%%%%%%%%%%%%%%%%%%%%%%%%%%%%%%%%%%%%%%%%%%%%%%%%%%%%%

\subsection{Mediated interactions in 1D}
%\label{Mediated interactions between static particles}
Two mobile bosonic impurities in a 1D Fermi gas were analysed using the Bethe ansatz and an effective model~\cite{huber2019medium}. 
A decrease in the energy was interpreted as an attractive mediated interaction between the impurities, which was shown to hold 
bound states in certain regimes.
The properties of two fermionic impurities interacting with a bath five fermions of another kind in a 1D harmonic trap were explored using a numerical approach for solving the time-dependent many-body Schr\"odinger equation. This method relies on  expanding the many-body wave function in terms of a time-dependent and variationally optimized basis.  
It permits to extract the interspecies and intraspecies correlation functions, 
the equilibrium features, and their dynamical properties~\cite{mistakidis2019repulsive}. The average 
distance between the two fermions in the trap were found to decrease with increasing impurity-bath interactions, which was 
interpreted as the presence of an attractive interaction mediated by the bath. Similar results were found 
indicative of an attractive mediated interaction for two 
bosonic impurities interacting with six fermions in a 1D harmonic trap~\cite{Mukherjee2020}.

The interaction between two static impurities separated by a distance $r$ in 1D  quantum liquids was 
explored following a path-integral formalism~\cite{Recati2005}.
For a non-interacting Fermi gas, the interaction between two impurities in the limit of weak fermion-impurity interaction is
quadratic in the impurity-fermion interaction $\gamma=g/v_F$
\begin{equation}
V(r) =-\gamma^2\frac{v_F}{2\pi r}\cos(2\pi p_F r),  
\end{equation}
where $v_F$ is the Fermi velocity. It is mediated the exchange of a particle-hole excitations, 
exhibits Friedel oscillations, and  is the 1D analogue of the RKKY interaction in Eq.~\eqref{RKKYrealaspace}. 
For strong impurity-fermion interaction, the mediated interaction  takes the form
\begin{equation}
V(r) =\frac{v_F}{2\pi r}\text{Re}\,\text{Li}_2(e^{2ip_Fr}), 
\end{equation}
where $\text{Li}_2$ is the di-logarithmic function. Remarkably, the mediated interaction is independent of the impurity-fermion interaction strength in this regime.  
When the two static impurities are immersed in an interacting Fermi gas forming a Luttinger liquid, the mediated interaction is proportional to $1/r$
for large distances~\cite{Recati2005}.  

{Interestingly, the theoretical framework used to study Fermi polarons has found direct applications in nuclear physics, such as cluster formation in neutron-rich matter~\cite{Tajima2024,Tajima2024b,TAJIMA2024138567}.}

\section{Quantum Gases: Boson-Mediated Interactions}
\label{quantumBosegases}
In this section, we turn to the case of interactions between impurities in BECs motivated by the rapid progress in ultracold atomic gas experiments.

\subsection{Mediated interactions in 3D}
As mentioned above, a very powerful feature of atomic gases is that  the interaction between the impurity atoms and the surrounding 
majority atoms can be tuned to be very strong using a Feshbach resonance. In order to analyse this regime, a non-perturbative diagrammatic theory for the quasiparticle 
interaction between Bose polarons  was developed in Ref.~\cite{camacho2018landau}. The theory is based on using an extended self-energy for the impurities in 
Eq.~\eqref{landausigma}, since it turns out to be necessary to go beyond the ladder approximation  to recover the second 
order result given by Eqs.~\eqref{RKKY} and \eqref{compressibilitybosons} for weak interactions. 
The resulting quasiparticle interaction is %shown diagrammatically in Fig.~\ref{FigBECmediatedStrong}(a) and is given by 
\begin{gather}
f(\mathbf p_1,\mathbf p_2)= \pm n_0Z_{\mathbf{p}_1}Z_{\mathbf{p}_2}\left[\mathcal{T}^2(p_1)G_{11}(p_1-p_2)+\right.\nonumber\\
\left.\mathcal{T}^2(p_2)G_{11}(p_2-p_1)+2\mathcal{T}(p_1)\mathcal{T}(p_2)G_{12}(p_2-p_1)\right].
\label{feff}
\end{gather}
Here, $n_0$ is the density of the BEC,  $Z_{\mathbf{p}}$ is the residue of a Bose polaron with momentum $\bf p$,  
$p_i=({\mathbf p}_i,\varepsilon_{\mathbf p_i})$ is the on-shell polaron momentum/energy, and $G_{11}$ and $G_{12}$ are the normal 
and anomalous Green's functions of the BEC~\cite{camacho2018landau}. Equation \eqref{feff} is a strong coupling extension of the second order 
result, which includes two-body impurity-boson correlations exactly. It 
 can be obtained basically by substituting the weak coupling scattering matrix ${\mathcal T}_v=2\pi a/m_r$ with the ladder expression 
$\mathcal{T}(p)={\mathcal T}_v/[1-{\mathcal T}_v\Pi(p)]$ with $\Pi(p)$ the boson-impurity pair propagator.

\begin{figure}
\includegraphics[width=0.99\columnwidth]{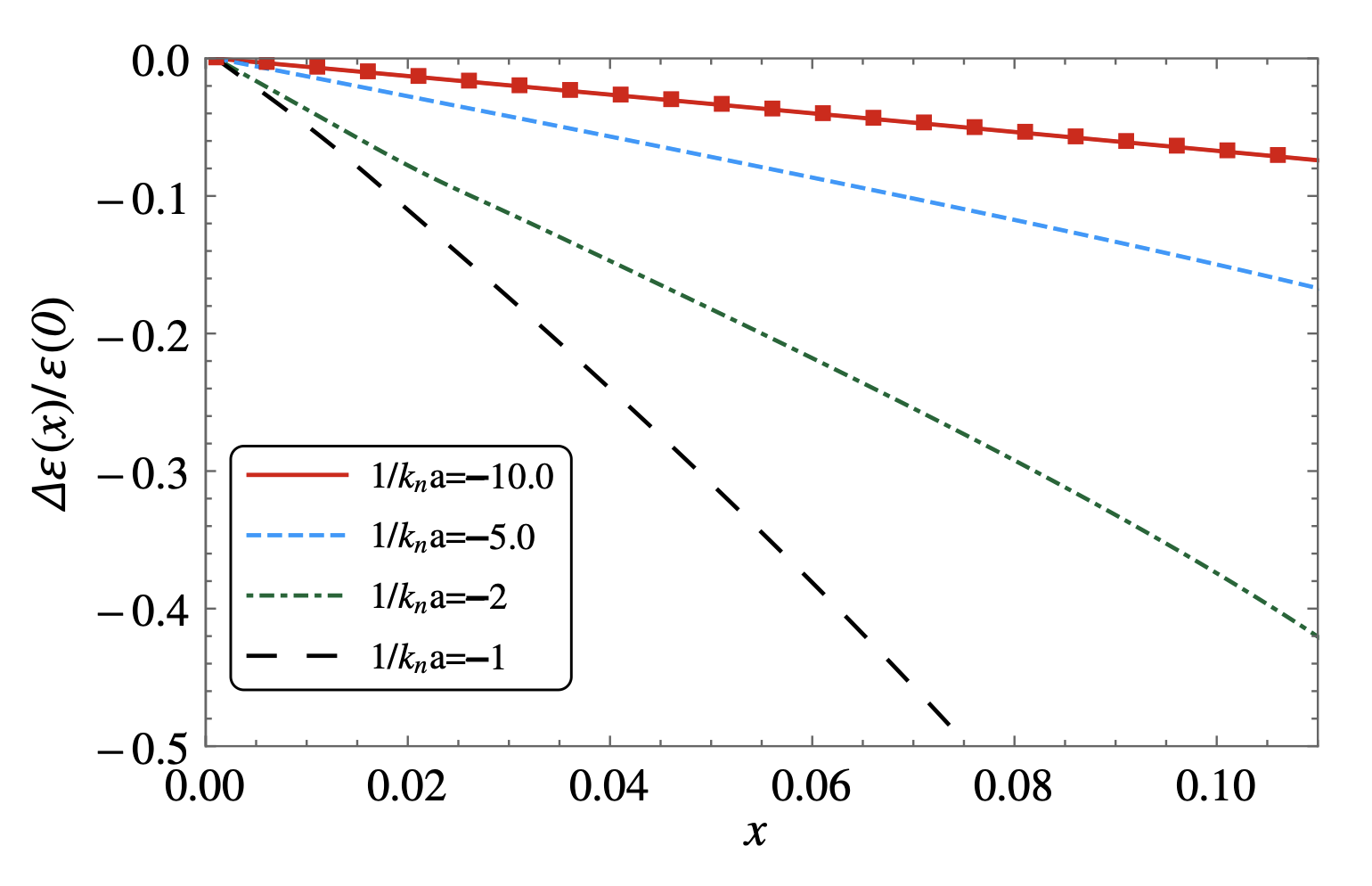}
\includegraphics[width=0.99\columnwidth]{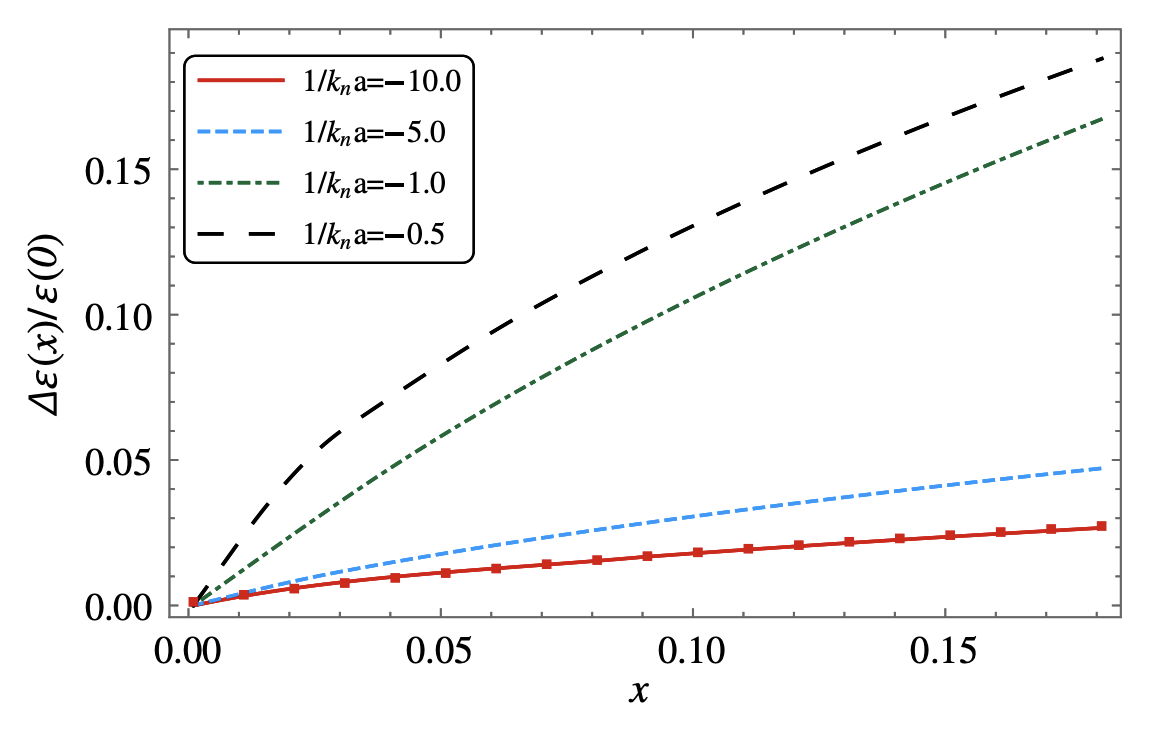}
\caption{Figures taken from Ref.~\cite{camacho2018landau}. %Top: Feynman diagram for the quasiparticle interaction. Black lines 
Top: The energy shift $\Delta\epsilon=\epsilon(n)-\epsilon(0)$ 
of a Bose polaron, where $\epsilon(n)$ is the polaron energy for 
concentration $n$ and $x=n/n_B$ with $n_B$ the concentration of the BEC. The mass ratio is $m/m_B=1$, the temperature  is
$T=0.4T_C$ with $T_C$ the critical temperature of the BEC, and the boson-boson scattering length is $k_na_B=0.2$. The energy shift is shown 
for various impurity-boson scattering lengths $k_na$ and the perturbative result $\Delta\epsilon=-ng^2/g_b$ from Eq.~\eqref{MediatedIntBEClowmomentum} 
is shown as red squares for $1/k_na=-10$.
Bottom: The energy shift of a Bose polaron for mass ratio $m/m_B=40/7$, $T=0$, $k_na_B=0.2$, and various impurity-boson scattering lengths.} 
\label{FigBECmediatedStrong}
\end{figure}

Using this approach, it was shown that the quasiparticle interaction between polarons can be strong close to impurity-boson resonance
with $1/k_na=0$, which leads  to significant energy 
shifts for the polaron energy. This is shown in Fig.~\ref{FigBECmediatedStrong}, where the polaron energy is plotted a function of its concentration for 
a bosonic impurity in a BEC with mass ratio $m/m_B=1$ (left) and a fermionic impurity with  
mass ratio $m/m_B=40/7$ (right) corresponding to  $^{40}$K impurity atoms in a $^7$Li BEC.
%%%%%%%%%%%%%%%%%%%%%%%%%%%%%%%%%%%%%%%%%%%%%%%%%%%%%%%%%%%%%%%%%%%%%%%%%%%%%%%%%%%%%%%%%
%%%%%%%%%%%%%%%%%%%%%%%%%%%%%%%%%%%%%%%%%%%%%%%%%%%%%%%%%%%%%%%%%%%%%%%%%%%%%%%%%%%%%%%%%
We see that the energy shift is negative for a bosonic impurity {(top)} corresponding to an attractive polaron-polaron interaction,
whereas it is positive 
for a fermionic impurity {(bottom)} corresponding to a repulsive interaction consistent  with 
the  results in Sec.~\ref{Bosepert}. 
Even though such energy shifts should be detectable, the quasiparticle interaction between Bose polarons remains to be observed, possibly due to the 
generally broad experimentally energy spectra. It has been shown that 
an alternative way to observe the BEC mediated interaction is via the sizable frequency shift it causes for the out-of-phase dipole mode of two Fermi gases separated by a BEC as compared to their in-phase mode~\cite{Suchet2017}

Using the interaction given Eq.~\eqref{feff} in an effective Schr\"odinger equation for two polarons, it was 
 predicted to be strong enough to support bound states, i.e.\ 
bi-polarons~\cite{Camacho2018}. The 
results for the binding energy of the bi-polaron were furthermore shown to compare well with Monte-Carlo calculations, 
and to recover exact results for the bound states of the Yukawa interaction given by Eq.~\eqref{VYukawa} for  weak 
interactions.  
The same approach was also used to predict  bound states between two Bose polarons both below and above the two-particle continuum 
in a 2D square lattice due to an attractive quasiparticle interaction~\cite{ding2023polarons}. Within the Fr\"ohlich model, valid for weak interactions, bipolarons were furthermore predicted using an all-coupling treatment  in Ref.~\cite{Casteels|2013}. The problem of a gas of diluite polarons  has been studied recently with perturbative a field-theory and a quantum–Monte Carlo (QMC) method in Ref.~\cite{ardila2022ultra}. {Squeezing and entanglement of distinguishable polarons have been studied in the context of two quantum Brownian particles coupled to a bosonic bath~\cite{charalambous2019two}.}

%%%%%%%%%%%
%%%%%%%%%%
%%%%%%%%%%%

\begin{figure}
\includegraphics[width=1
\columnwidth]{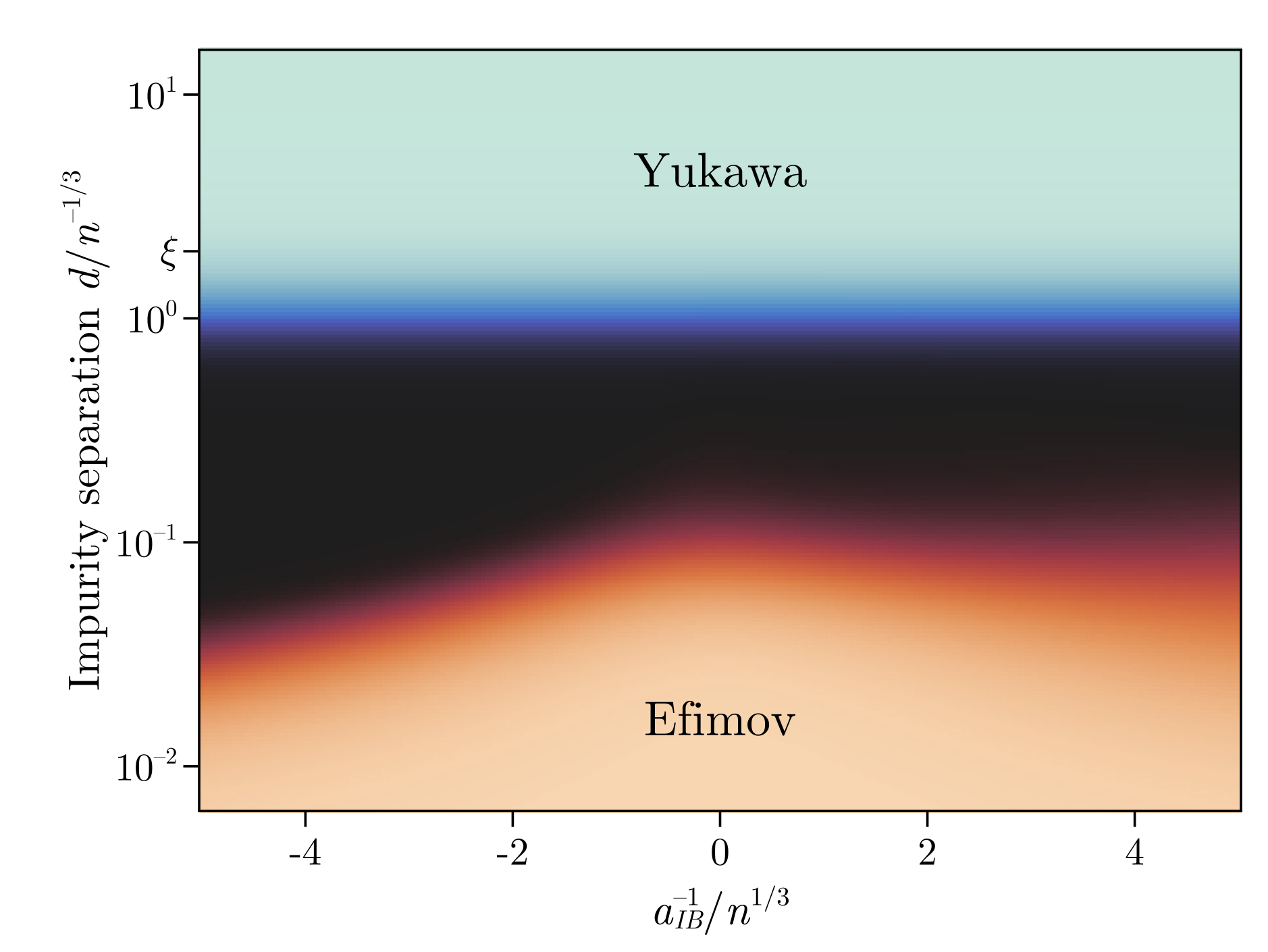}
\caption{Figure  from Ref.~\cite{drescher2023medium}. Sketch of the regimes of Efimov and Yukawa scaling of the mediated interaction between two static impurities in a BEC as a function of the inverse of the impurity-boson scattering length $a_{IB}$ and the distance between the impurities. Here, $n$ is the density of the bosonic bath.    
} 
\label{FigEnss}
\end{figure}

The induced interaction between two static impurities was calculated within a Gross-Pitaevskii  theory (GPT) and its nonlocal extension (NLGPT)~\cite{drescher2023medium}. In this approach, the induced interaction between the impurities is determined by comparing the energy of two impurities $E_d$ fixed at a distance $d$ with the energy of two individual impurities infinitely far apart $E_\infty$, such that $V_{\text{ind}}=E_d-E_\infty$.
For short distances compared to the interparticle spacing, it was shown that the interaction is dominated by the three-body physics of a single boson bound between the two static impurities, giving rise to an Efimov-type $V_{\text{ind}}\propto 1/r^2$ scaling. For larger distances of the order of the coherence length of the BEC, the interaction has the Yukawa form 
Eq.\eqref{VYukawa}. This crossover between Efimov and Yukawa scalings of the mediated interaction was found for every impurity-boson interaction strength contrary to an earlier analysis based on a variational calculation restricted to a single Bogoliubov mode,
which found a Yukawa potential for weak interactions and an Efimov interaction for resonant impurity-boson interaction~\cite{naidon2018two}.
The different regimes of the  interaction between two static impurities mediated by a BEC of density $n$ are shown in Fig.~\ref{FigEnss} 
as a function of the impurity separation $d$ and the boson-impurity scattering length $a_{ib}$. 
In Ref.~\cite{naidon2018two}, the interplay between mediated interactions and three-body Efimov states were furthermore explored. 
 
In the same spirit,  the interaction  between two static impurities mediated by a general superfluid with a linear  phonon 
dispersion was calculated using effective field theory~\cite{fujii2022universal}. Assuming weak and short-range impurity-superfluid interactions, it was shown that the mediated interaction is dominated by the exchange of two phonons for  large distances $r\gg \xi$, resulting in a $1/r^7$ scaling instead of the Yukawa interaction mediated by one phonon exchange dominant 
for shorter distances. This result holds in general for a superfluid with a low energy 
linear phonon dispersion with the sound velocity determining the strength of the interaction.

Considering again two static impurities in a BEC, the effects of boson-boson interactions were shown to be important for the mediated interaction and 
bi-polaron formation for strong impurity-boson interactions~\cite{jager2022effect}.
The interaction between two static impurities in a BEC was also analysed using a path integral approach~\cite{Panochko2022}.

%%%%%%%%%
%%%%%%%%
%%%%%%%%%

Induced interactions are not limited to the case where the impurity-boson interaction is short-ranged. For instance, progress creating hybrid ion-atom systems~\cite{Grier2009,Harter2012,Kleinbach2018,Tomza2019,Dieterle2021} has stimulated  much attention
to charged polarons~\cite{casteels2011polaronic,astrakharchik2021ionic,Christensen2021,Christensen2022,pessoa2024fermi} and their 
interactions. The interaction between two static ions immersed in a neutral atomic BEC or Fermi gas was 
considered in Ref.~\cite{ding2022mediated}. {For the weakly interacting case, it was shown that the interaction between two ions mediated by a BEC, exhibits a 
Yukawa form for intermediate distances, and a  $1/r^4$ power-law behaviour for large distances, reminiscent of the ion-atom interactions. }{This Yukawa interaction for long distances arises also for the mediated interaction between Rydberg impurities~\cite{RydbergWang}.}
For strong atom-ion interactions, a calculation similar to that  leading to Eq.~\ref{feff} shows  that for large distances 
the mediated interaction is proportional to $ \mathcal T(0,0;0)/r^4$ where  $\mathcal T(0,0;0)$ is the zero energy/momentum 
atom-ion scattering matrix. For two static ions interacting weakly with  a  Fermi gas, the mediated interaction is given by a power-law for large density and by a RKKY form for low density. 
Using  Monte Carlo simulations, the interaction between  two static impurities in a weakly interacting BEC was studied~\cite{astrakharchik2023many}.
The ions were found to strongly perturb the BEC in their surroundings and  different regimes related to the formation of bi-polarons were identified.

\subsection{Mediated interactions in 1D}
The interaction between  two mobile impurities mediated by bosons in a 1D harmonic trap has been studied using a numerical 
solution to the  time-dependent many-body Schr\"odinger equation. 
By calculating the average distance between the impurities as well as non-equilibrium dynamics,  
signatures of attractive and repulsive 
mediated interactions  were identified and bound bi-polaron states  predicted~\cite{Mistakidis2020,mistakidis2019correlated,mistakidis2020many,theel2023crossover}.

\begin{figure}
\includegraphics[width=0.99\columnwidth]{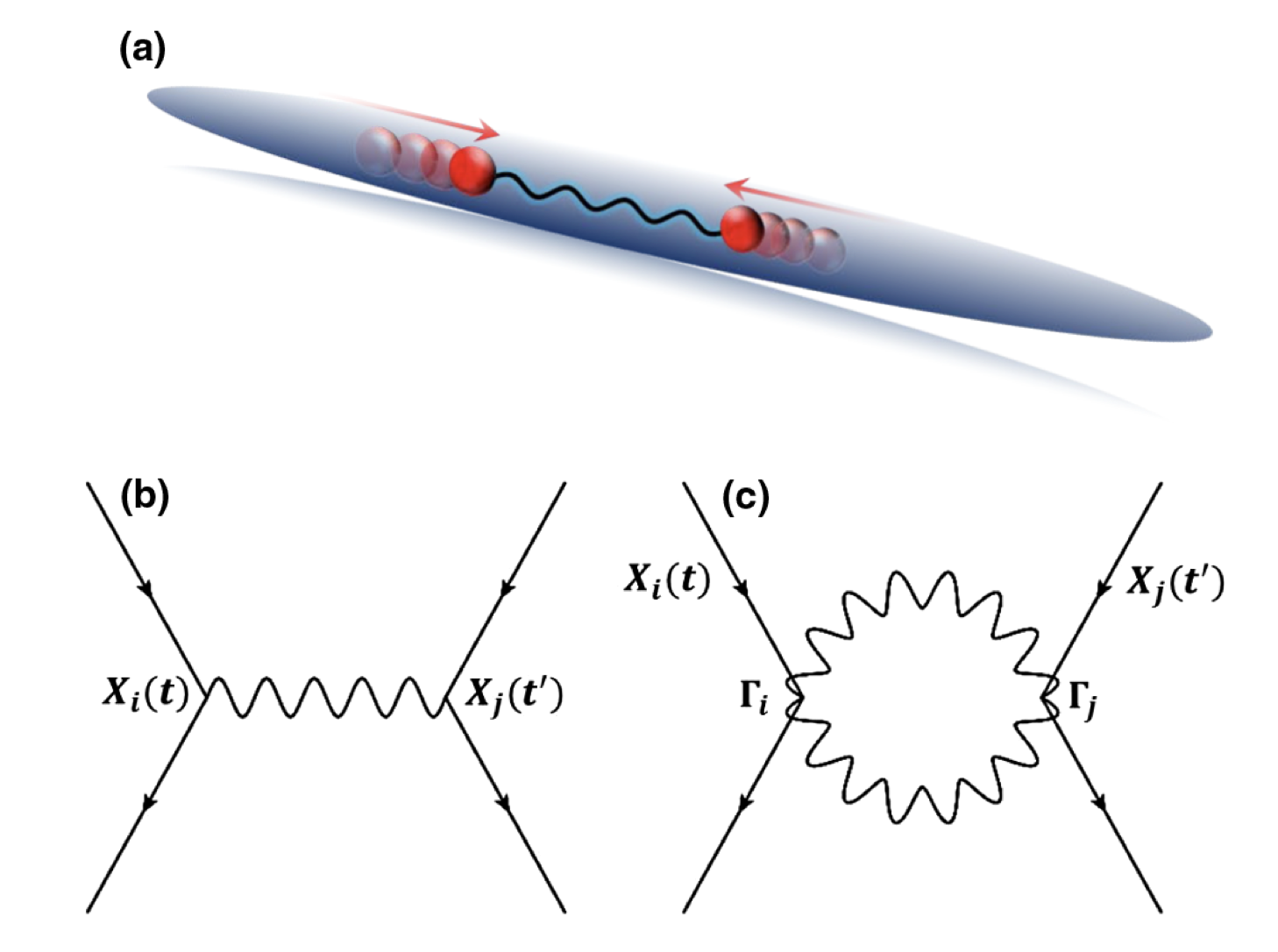}
\caption{Figure  from Ref.~\cite{Schecter2014}. (a) The interaction between two static impurities mediated by a 1D Bose gas. (b) Single-phonon exchange responsible for the exponential decay of the interaction at short distances. {Indeed, this process corresponds to the one discussed in Fig.~\ref{Fig1}(c). } (c) Two-phonon exchange giving rise to a long-range Casimir-like interaction between the impurities. The solid lines correspond to the impurity propagator while the wavy line illustrate the phonon modes.} 
\label{FigCasimir}
\end{figure}

The interaction between two static impurities separated by a distance $r$ 
in a 1D repulsively interacting Bose gas has been studied by numerous authors  and is 
illustrated in Fig.~\ref{FigCasimir}(a). For short distances $r\lesssim \xi$ compared
to the healing length of the bosons, the dressing clouds of the two impurities overlap giving rise to an 
interaction~\cite{Klein2005,Recati2005,Dehkharghani2018}
\begin{equation}\label{1DBoseShort}
V(r)=-\sqrt{\frac{n_0m_B}{g_b}}g^2e^{-2r/\xi}.
\end{equation}
This  can be derived  from a Fourier transform of the  1D version of Eq.~\ref{compressibilitybosons} and 
corresponds to the exchange of a single phonon between the impurities as illustrated in Fig.~\ref{FigCasimir} (b). {This diagram corresponds to the one described in detail using second-order perturbation theory in Fig.~\ref{Fig1}(c), where impurities exchange a Bogoliubov mode from the condensate. }
For large distances $r\gg \xi$ where the two dressing clouds do not overlap, the 
interaction between the impurities is due to quantum fluctuations in close analogy with the Casimir effect as already discussed in Sec.~\ref{fermimediated}. This gives rise to the interaction~\cite{Schecter2014,Pavlov2019,Reichert2019,Reichert2019b} 
\begin{equation}\label{1DBoseLong}
 V(r)=-\frac{g^2 m_B\xi^3}{32\pi r^3},
\end{equation}
exhibiting the same $1/r^3$ scaling as the Casimir force. It is caused by the exchange of two phonons as 
illustrated in Fig.~\ref{FigCasimir} (c).
Further studies concerning polaron interactions and bipolarons in 1D emphasised the  deformation of the superfluid by the impurities~\cite{Will2021,Brauneis2021}, the presence of lattice 1D bipolarons~\cite{Isaule2024}, {and bipolarons maximally entangled appearing in the form of Bell states~\cite{anhtai2024engineeringimpuritybellstates}.}

\subsection{Photon-mediated interactions}
\label{QOL}
With the strong coupling of atoms to photons  in optical cavities, a new way to engineer hybrid light-matter 
states has emerged~\cite{carusotto_quantum_2013,mivehvar2021cavity}. Quantum optical lattices refer to ultracold quantum gases confined to optical lattices embedded in high-finesse cavities, where the quantum nature of the photons and the atoms have to be
treated on the same footing~\cite{maschler2008ultracold}. As we will now describe, the photons mediate a long-range interaction between the atoms, which can lead to a range of 
interesting phenomena~\cite{Park2022}. 
In this subsection, we first discuss how derive this interaction following a diagrammatic  procedure as  discussed in Sec.~\ref{Green} in this  Perspective. We then connect this 
to a more conventional derivation using the the Heisenberg equations of motion~\cite{mivehvar2021cavity}.

\begin{figure}
\includegraphics[width=0.99\columnwidth]{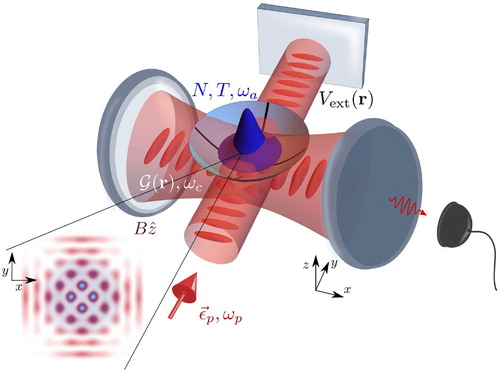}
\caption{Figure  from Ref.~\cite{mivehvar2021cavity}: Cartoon of $N$ atoms loaded into an external potential $V_{\text{ext}}$ and sandwiched between two mirrors (grey plates).  Cavity photons mediate long-range atomic interactions between atoms in its ground state. The interplay between the tunneling dynamics, on-site atomic interactions, and induced infinite-range interactions between the atoms leads to a rich phase diagram.} 
\label{FigQC}
\end{figure}

\subsubsection{General Theoretical framework}
To establish a general model that describes the photon mediated interaction between atoms, we consider the scheme of Fig.~\ref{FigQC}. There, an ensemble of $N$ two-level atoms having a transition frequency $\omega_a$ between its internal states $\{ |g \rangle, | e \rangle \}$ is inside a single-mode linear cavity with wavelength $\lambda_c$ and resonance frequency $\omega_c=2 \pi c/ \lambda_c=ck_c$, with $c$ the speed of light. The atoms are strongly coupled to the cavity photons 
with an amplitude $\mathcal {G}(\mathbf{r})= \mathcal{G}_0 \cos{k_c x}$, where the oscillatory factor reflects the amplitude of the cavity mode. They are also subjected to a transverse pump laser with polarization $\mathbf{\epsilon}_p$ and frequency $\omega_p$ close to the cavity resonance frequency. The coupling between this transverse pump laser and the atoms is described  by the position dependent Rabi frequency $\Omega(\mathbf{r} )= \Omega _0 \cos{k_c y}$. In the rotating frame of the pump laser, the system is described by $\hat H =\hat{H}_0 + \hat{H}_{\text{int}}$, with $\hat{H}_0$ given by
 \begin{gather}
\hat{H}_0 =  \sum_{n= g,e} \int {\hat \psi}_n^\dagger(\mathbf{r}) \left[ -\frac{\hbar^2}{2m} \nabla^2 + V_{\text{ext}} (\mathbf{r})\right ] {\hat \psi}_n (\mathbf{r}) \text{d}\mathbf{r}  \nonumber\\ \nonumber - \hbar \Delta_c { \hat a }^\dagger { \hat a } - \hbar \Delta_a \int \hat{\psi}_e^\dagger(\mathbf{r}) \hat{\psi}_e(\mathbf{r}) \text{d}\mathbf{r}  \\ 
+\hbar \int \{ \hat{\psi}_e^\dagger(\mathbf{r}) \left[ \Omega(\mathbf{r}) + \mathcal{G}(\mathbf{r}) \hat{a} \right] \hat{\psi}_g(\mathbf{r})+ \text{H.c.} \} \text{d}\mathbf{r}.
\end{gather}
Here ${\hat \psi}_n (\mathbf{r})= {\hat \psi}_n (\mathbf{r},t)$ and $\hat{a}=\hat{a}(t)$ are the slowly-varying  atomic
and photonic annihilation field operators, $V_{\text{ext}}(\mathbf{r})$ is an external trapping potential for the atoms, $\Delta_a=\omega_p-\omega_a$ is the difference between the pump-laser frequency and the atomic transition frequency, and $\Delta_c= \omega_p- \omega_c$ the relative frequency between the pump-laser and the cavity frequencies. The atomic two-body interactions are represented by $\hat{H}_{\text{int}}$, which in the limit of low temperatures and low density are described by means of a contact interaction
\begin{gather}
\hat{H}_{\text{int}}= \frac{1}{2}\sum_{n= g,e} g_{nn} \int {\hat \psi}_n^\dagger(\mathbf{r}) {\hat \psi}_n^\dagger(\mathbf{r}) {\hat \psi}_n(\mathbf{r}) {\hat \psi}_n(\mathbf{r})  \nonumber  \\+ g_{eg} \int {\hat \psi}_e^\dagger(\mathbf{r}) {\hat \psi}_g^\dagger(\mathbf{r}) {\hat \psi}_g(\mathbf{r}) {\hat \psi}_e(\mathbf{r}).
\end{gather}
The coefficients $g_{ee}$, $g_{gg}$, and $g_{eg}$ can be written in terms of the $s$-wave scattering lengths $a_{nn'}$ as $g_{nn'}= 4\pi \hbar^2 a_{nn'}/m$. Here it must be pointed out that for fermionic atoms, owing to Pauli exclusion principle no interactions take place among same internal states. The scattering lengths $a_{nn'}$ can be varied externally through Feshbach resonances as a magnetic field is tuned.

\begin{figure}
\includegraphics[width=0.99\columnwidth]{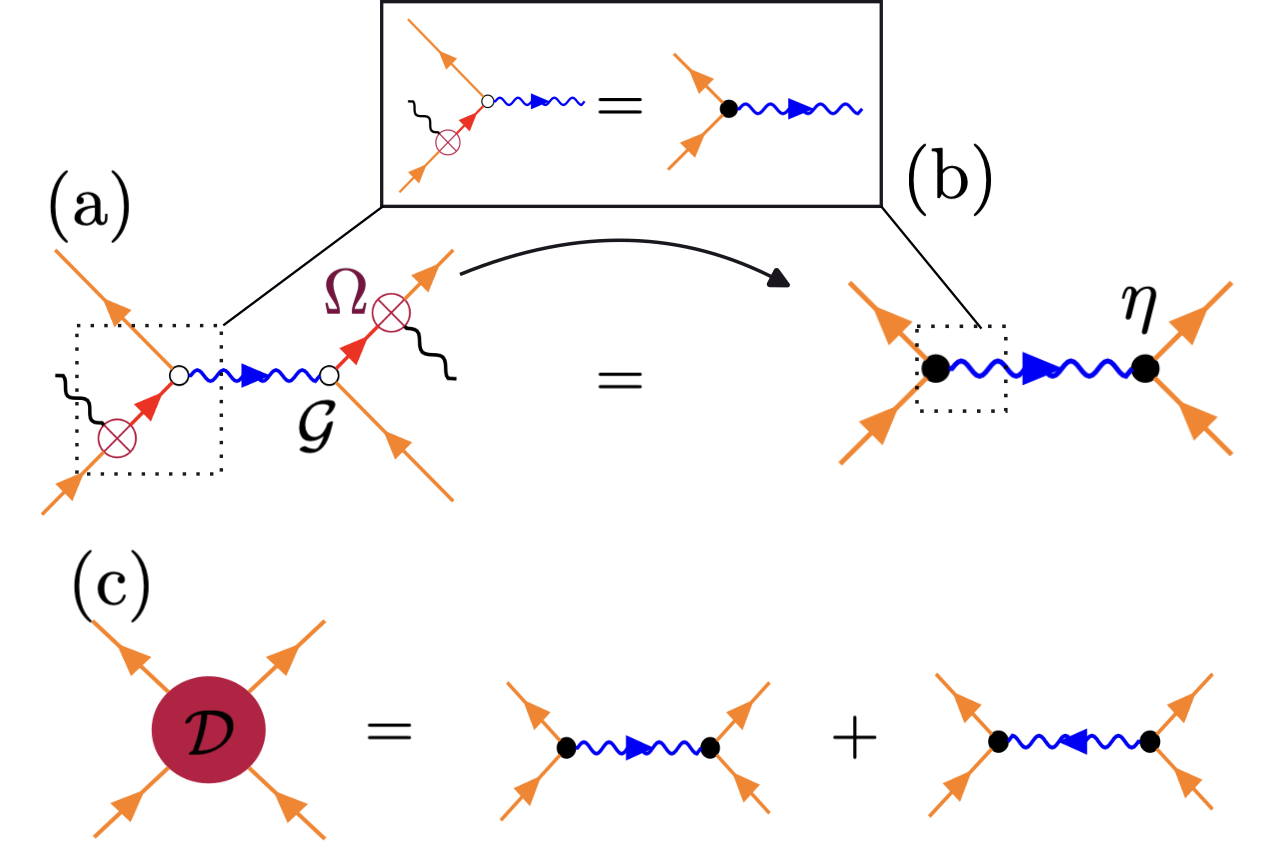}
\caption{Diagrammatic representation of the photon mediated interaction. {(a) Mediated interaction to  lowest order in $\Delta_a$ involving the exchange of a single photon. The solid orange lines represent the ground state propagator, the red lines depict the excited state propagator. As illustrated in the inset, the photon exchange process arises from a second-order process where an atom from the ground state is excited via a classical pump field (crossed dot with the black wiggly line), which then emits a photon (wavy blue line), which is absorbed in the reversed process from the inset. (c) The effective interaction between ground state atoms is denoted by $\mathcal{D}.$
 }
%between ground state atoms to the lowest order in $\Delta_a$, which involves the exchange of a single photon. The solid yellow/red lines depict the atomic ground/excited state Green's functions, 
%the crossed dot with the black wiggly line corresponds to the %classical pump field, 
%the white dot corresponds to the coupling to the cavity photons, and the wavy blue line is the photon propagator. 
} 
\label{FigPhotonDiagram}
\end{figure}
{\bf Diagrammatic Formulation.- } Following the Green's function formalism discussed in Sec.~\ref{Green}, we now construct the relevant diagrams for the photon-mediated interaction among ground state atoms. Such a 
diagrammatic approach was recently used to analyse the Sachdev--Ye--Kitaev model in a quantum cavity~\cite{uhrich2023cavity}.

First, our main assumption is that the pump and cavity frequencies are far detuned from the excited state such that $\Delta_a$ is large. This is the so-called dispersive regime where the population of the excited 
state is small and one can write their Green's function as $G_e(k)\simeq1/\Delta_a$. The leading diagram in $1/\Delta_a$ for the photon-mediated interaction between the ground state atoms 
can then  be drawn as in Fig.~\ref{FigPhotonDiagram}(a), which intuitively can be interpreted from left to right as: a ground state atom (orange line) is excited to the excited state (red line) by the classical 
pump field $\Omega$ (red cross with {the black wiggly} line). The excited state then decays to the ground state again by emitting a photon into cavity mode  {(wavy blue line)}.  The cavity 
photon is subsequently absorbed by a second ground state atom, which is then excited. This atom finally decays back into the ground state atom via stimulated emission into the classical field. This completes the mediated interaction, as two ground state atoms effectively exchange a photon. As illustrated in Fig.~\ref{FigPhotonDiagram}(b), the processes highlighted by the box can be summarized into an effective interaction $\eta({\mathbf r})= \Omega({\mathbf r}) \mathcal{G}({\mathbf r})/\Delta_a$ between a ground state atom and a photon. Thus, as for the other mediated interactions discussed in this perspective, 
the photon mediated interaction between the ground state atoms is quadratic in $\eta$. Finally, assuming that the dynamics of the cavity photon can be ignored when $\Delta_c$ is large so that its Green's 
function is {$G_c(k)\simeq1/\Delta_c$} and taking into account the two processes in Fig.~\ref{FigPhotonDiagram}(c) gives 
\begin{gather}
\mathcal{D}(\mathbf r, \mathbf r')=\frac{2 \Omega_0^2 \mathcal{G}_{0}^2}{\Delta_a^2\Delta_c} \cos(k_c x') \cos( k_c y') \cos( k_c x) \cos (k_c y),
\label{eff_a-a_Bose}
\end{gather}
for the photon mediated interaction between the ground state atoms. {As described below, one  reaches this expression for $\mathcal{D}(\mathbf r, \mathbf r')$, following a standard procedure for the Heisenberg equations of motion for the atom and photon field operators.}
%Physically, since the coupling between the photons and the ground state atoms involves an excited state, the mediated interaction should be at least quadratic in $1/\Delta_a$. 
Note that this interaction is infinite range reflecting that the cavity mode is assumed to cover the whole system. 
%\rosario{This mediated interaction corresponds to the Feynman in Fig.~\ref{FigPhotonDiagram}, quadratic in the excited state propagator $1/\Delta_a$, the coupling $\Omega$ as well as quadratic on $\mathcal G$ as discussed previously. }
The photon-mediated interactions can be derived alternatively using a path-integral formalism as in Sec.~\ref{pathI}. This approach has been employed for the study of supersolid phases of  bosonic atoms inside two crossed optical cavities \cite{lang2017collective,piazza2013bose}.

{\bf Heisenberg equations of motion.-} 
It is illuminating to present yet another derivation of the photon mediated interaction using the Heisenberg equations of motion, which is common in the field of quantum optics.
To do this, we closely follow the approach in Ref.~\cite{mivehvar2021cavity}.

Let us now start from the Heisenberg equations for the  atomic  and photonic field operators 
\begin{gather}
i\hbar \frac{\partial {\hat \psi}_e(\mathbf{r})}{\partial t} = \left[-\frac{\hbar^2}{2m} \nabla^2 + V_{\text{ext}} (\mathbf{r}) -\hbar \Delta_a + g_{ee} {\hat \psi}_e^\dagger(\mathbf{r}) {\hat \psi}_e (\mathbf{r})  \right.+\nonumber\\ 
\left.
 + g_{eg} {\hat \psi}_g^\dagger(\mathbf{r}) {\hat \psi}_g(\mathbf{r}) \right] {\hat \psi}_e(\mathbf{r}) + \hbar\left[\Omega(\mathbf{r}) + \mathcal{G}(\mathbf{r}) \hat{a} \right]  {\hat \psi}_g(\mathbf{r}), 
\end{gather}
and
 \begin{gather}
 i\hbar \frac{\partial {\hat \psi}_g(\mathbf{r})}{\partial t} = \left[-\frac{\hbar^2}{2m} \nabla^2 + V_{\text{ext}} (\mathbf{r}) + g_{gg} {\hat \psi}_g^\dagger(\mathbf{r}) {\hat \psi}_g (\mathbf{r}) \right.\nonumber\\ \left.
 + g_{eg} {{\hat \psi}_e^\dagger(\mathbf{r}) {\hat \psi}_e(\mathbf{r})} \right] {\hat \psi}_g(\mathbf{r}) + \hbar\left[\Omega(\mathbf{r}) + \mathcal{G}(\mathbf{r}) \hat{a}^\dagger \right]  {\hat \psi}_e(\mathbf{r}) 
 \label{Derivativepsig}
\end{gather}
\begin{gather}
i\hbar \frac{\partial \hat{a}}{\partial t} = -\hbar \Delta_c \hat{a} + \hbar \int \mathcal{G}(\mathbf{r})  {\hat \psi}_g^\dagger(\mathbf{r}) {\hat \psi}_e (\mathbf{r}) \text{d}\mathbf{r}.\label{Derivativea}
\end{gather}
We now use that in the dispersive regime where the pump and cavity frequencies are far detuned from the atomic transition frequency,  the fastest time scale is $1/\Delta_a$. This means  that the excited state is minimally populated and therefore has only negligible effects on the  dynamics. Using a procedure known as adiabatic elimination, the steady state of the atomic $|e\rangle$ field operator  becomes $\hat{\psi}_{e,ss} (\mathbf{r}) \approx \frac{1}{\Delta_a} [\Omega(\mathbf{r}) + \mathcal{G}(\mathbf{r}) {\hat a}] \hat{\psi}_g (\mathbf{r})$. After substituting this and 
${\hat \psi}_e^\dagger(\mathbf{r}) {\hat \psi}_e(\mathbf{r}) \propto 1/\Delta_a^2 $ into the equation of motion for ${\hat \psi}_g(\mathbf{r})$, one arrives at 
\begin{gather}
i\hbar \frac{\partial {\hat \psi}(\mathbf{r})}{\partial t} =%-\frac{g^2}{V}
\left[-\frac{\hbar^2 \nabla^2}{4m} + V_{\text{ext}}(\mathbf{r}) + \hbar V(\mathbf{r}) + \hbar U(\mathbf {r}) \hat{a}^{\dagger} \hat{a} \right.\nonumber\\ \left.
+ \hbar \eta ({\mathbf r}) (\hat{a}^\dagger + \hat{a}) +  g_0 \hat{n}(\mathbf{r}) \right] \hat{\Psi}(\mathbf{r}), 
\label{fieldphi}
\end{gather}
and
\begin{gather}
i\hbar \frac{\partial \hat{a}}{\partial t} = -\hbar \left( \Delta_c - \int U({\mathbf r}) \hat{n}({\mathbf r}) d{\mathbf r} \right)\hat{a} + \int \eta({\mathbf r}) \hat{n}({\mathbf r}) d{\mathbf r}.
\label{field_ph}
\end{gather}
Here, we have made the identifications, ${\partial_t {\hat \psi}_g(\mathbf{r})=\partial_t {\hat \psi}}(\mathbf{r})$, $V({\mathbf r})=\Omega^2({\mathbf r})/\Delta_a$, $U({\mathbf r})=\mathcal{G}({\mathbf r})/\Delta_a$, 
%$\eta({\mathbf r})= \Omega({\mathbf r}) \mathcal{G}({\mathbf r})/\Delta_a$, 
$\hat{n}(\mathbf{r})={\hat \psi}^\dagger_g(\mathbf{r}){\hat \psi}_g(\mathbf{r})$, and $g_{gg}=g_0$.
%with the corresponding amplitudes being $V_0= \Omega_0/\Delta_a$,  $U_0= \mathcal{G}_0/\Delta_a$, and $\eta_0= \Omega_0 \mathcal{G}_0/\Delta_a$. 
The above equations establish the coupled dynamics between matter and light. %\arturo{After substituting the stationary state for the photon field operator in $\partial_t {\hat \psi}$, the resulting effective interaction for the reduced atom system reads
%\begin{gather}
%V_{\text{ind}}= \int \hat{\psi}^{\dagger}(\mathbf{r}) \left[ -\frac{\hbar^2}{4m} \nabla^2 + V_{\text{ext}}(\mathbf{r})+\hbar V(\mathbf{r})\right.+\\ \left.\nonumber
%+ \frac{1}{2} g_0 n(\mathbf{r}) \right ] \hat{\psi}(\mathbf{r}) d\mathbf{r} + \int \int \mathcal{D}(\mathbf {r},\mathbf{r'}) n(\mathbf{r}) n(\mathbf{r'}) d\mathbf{r} d\mathbf{r'} 
%\end{gather}
%where $\mathcal{D}(\mathbf {r},\mathbf{r'})$ is the amplitude of the cavity-mediated infinite-ranged density-density interaction given in \ref{eff_a-a_Bose}.}

\begin{figure}
\includegraphics[width=0.99\columnwidth]{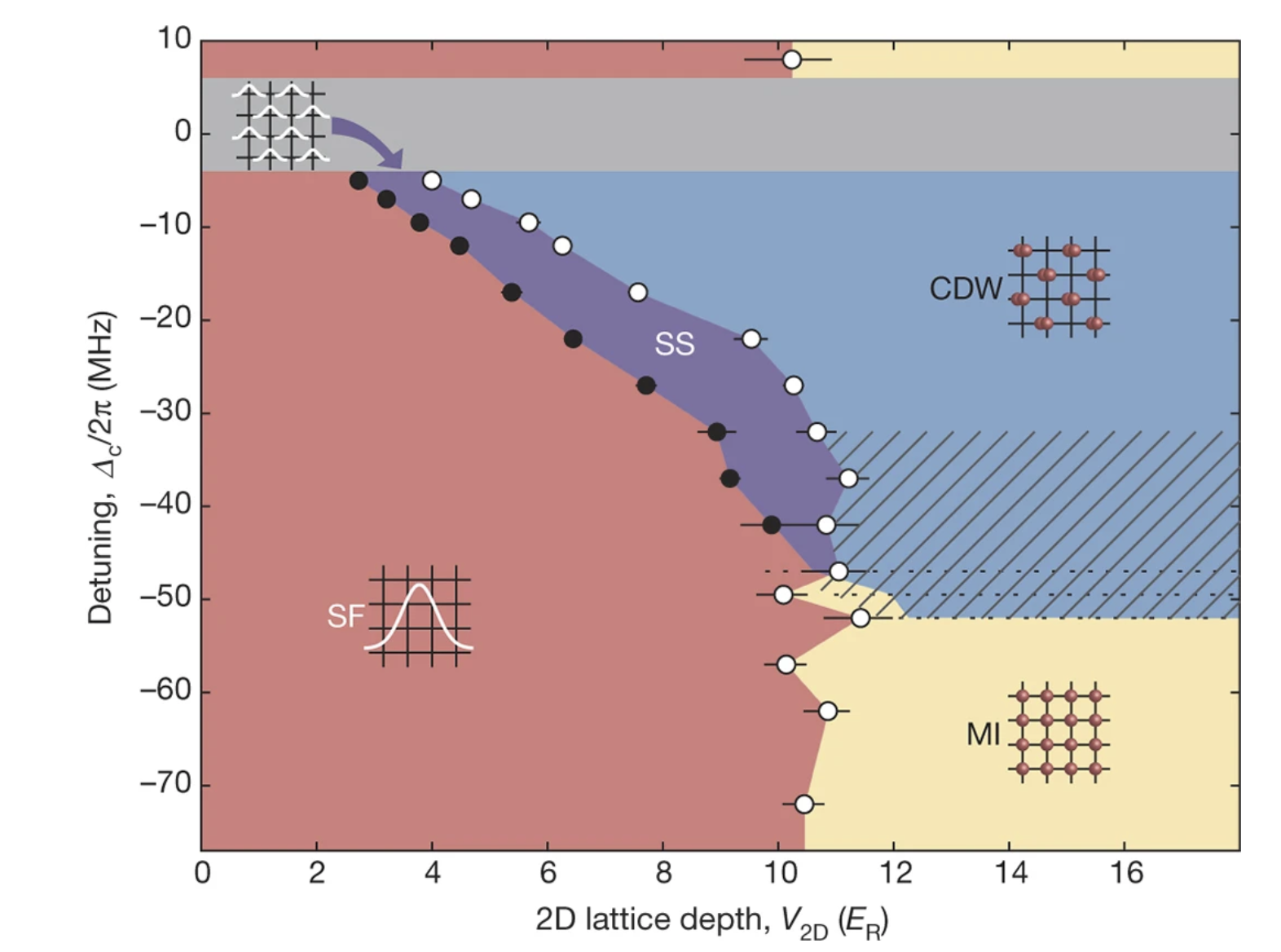}
\caption{Figure  from Ref.~\cite{Landig}: Experimental observation of cavity-induced many-body phases. Superfluid phase (red), Mott insulator (yellow), charge density wave (blue) and supersolidity (violet).
} 
\label{Esslinger2}
\end{figure}

We now consider the case in which the cavity field dynamics evolves much faster than the atom center of mass motion, which is equivalent to the assumption that $\Delta_c$ is large used in the diagrammatic derivation.
Then, the cavity field follows adiabatically the movement of the atoms, as in the Born-Oppenheimer approximation scheme of electrons in molecules. Using the steady-state solution to the photonic field operator $\partial_{t} \hat{a}=0.$ {For large detunings, this gives the photon field $\hat{a} \propto \frac{1}{\Delta_c}\int \eta({\mathbf{r}}) \hat{n}({\mathbf{r}}) d{\mathbf{r}}$, which is substituted into Eq.~\ref{fieldphi} and results in exactly the  same mediated interaction as the one derived diagrammatically
\begin{gather}
\hat{V}_{\text{med}}= \int \int \mathcal{D}(\mathbf {r},\mathbf{r'}) \hat{n}(\mathbf{r}) \hat{n}(\mathbf{r'}) d\mathbf{r} d\mathbf{r'}, 
\end{gather}
with $\mathcal{D}(\mathbf {r},\mathbf{r'})$ given by Eq.~\ref{eff_a-a_Bose}.
}
\subsubsection{Experiments and theoretical predictions}
The long range photon mediated interaction has given rise to a number of groundbreaking experiments including  the observation of the Dicke phase transition in a BEC coupled with cavity 
modes~\cite{BaumannK,BaumannK2,Brennecke,Klinder}. 
In pioneering experiments, four different quantum phases in a bosonic gas of $^{87}$Rb atoms in  layers of 2D 
square optical lattices in an optical cavity were observed, which were driven by the competition between the long range photon mediated interaction and a 
short range atomic interaction~\cite{Deng,Jun-RuLi,Helson2023,Landig,leonard2017supersolid}. 
As summarized in  Fig.~\ref{Esslinger2}, superfluid, supersolid, Mott insulating, and charge density wave phases were found in square lattices. 
In particular, the supersolid state is associated with the spontaneous breaking of two symmetries: the $U(1)$ symmetry related to the phase of the condensate wave function giving rise to superfluidity, and translational symmetry that produces the lattice structure.
The use of ring cavities has enabled the possibility of realizing other states and self-ordered phases.  \cite{Boddeda2016,Firstenberg_2016,Yang:22,Ostermann20,Suarez22,Hao19,JSheng,Vaneecloo}.

Other interesting results include photon-mediated atomic interactions in  multimode cavity QED ~\cite{Vaidya2018}, Higgs and Goldstone modes in a supersolid quantum gas~\cite{leonard2017monitoring}, spinor BEC self-ordering in a cavity~\cite{Kroeze2018}, and the formation of a spin textures in a quantum gas coupled to a cavity~\cite{Landini2018}. Also, 
disorder driven density and spin self-ordering of a Bose-Einstein condensate as well as  the  realisation of the Sachdev-Ye-Kitaev model by atoms in an optical cavity have been predicted~\cite{Mivehva2017,uhrich2023cavity} and random spin models with atoms in a high-finesse cavity~\cite{sauerwein2023engineering}. Quantum simulation of 
condensed matter phases with a long range photon mediated interaction is a rich topic~\cite{Caballero2016,Schafer2020} 
including a photon-mediated Peierls transition, emergent quasicrystalline symmetries \cite{Mivehvar2019}, multimode BECs \cite{Karpov2022}, supersolid phases \cite{Zhang2018}, spin entanglement and magnetic competition in spinor quantum optical lattices~\cite{Lozano2022},  bond order~\cite{caballero2016bond} as well as other interesting 
predictions~\cite{Grankin_2014,Caballero2015,Caballero2016,caballero2016bond,TitasChanda,Mivehva2017,PSchlawin2019,Camacho2017,Rylands2020,Karpov2022,Zhang2018,Lozano2022}. 
Reviews of cavity QED with quantum gases with varying detail can be found in Refs.~\cite{maschler2008ultracold,mekhov2012quantum,Ritsch2013,mivehvar2021cavity}.

\section{Two-dimensional semi-conductors}
\label{vanderWaals}
Novel semi-conductor materials have emerged as a platform in which new and interesting 
quantum many-body phases can been realized \cite{kennes2021moire}. 
This includes the realization of  strongly correlated states 
such as Wigner crystals \cite{Cirac2020,smolenski2021signatures}, Mott insulators \cite{Dubin2022,Wehling2023,Yuanbo2024}, conventional and unconventional  superconductivity~\cite{Jarillo2018,Guinea2022,YanfengGe2023}, 
as well as  new forms of hybrid light-matter 
states \cite{Takemura2014,Sidler2016,zhang2021van,zhao2023exciton,Jomaso2024}. Bose-Fermi mixtures can be created by means of electrons interacting with excitons or exciton-polaritons, which to a good approximation are bosonic due to their large binding energy in these 2D materials. 
In this section, we discuss the recent experimental and theoretical progress regarding mediated interactions between quasiparticles in these materials~\cite{Tan2020,tan2022bose,emmanuele2020highly,song2022attractive,muir2022interactions}.

\subsection{Fermi polaron-polariton interactions}\label{vanderWaalspolaron}
The experimental realization of highly imbalanced Bose-Fermi mixtures in 2D semiconductors allowed for the first observation of Fermi polaron-polaritons~\cite{Sidler2016}. In these experiments,  a two-dimensional electron gas (2DEG) plays the role of the fermionic bath, and excitons  play the role of the impurity particles. The 
excitons can furthermore be hybridised with cavity photons in which case the resulting quasiparticles are 
referred to as polaron-polaritons.

\begin{figure}
\includegraphics[width=0.85\columnwidth]{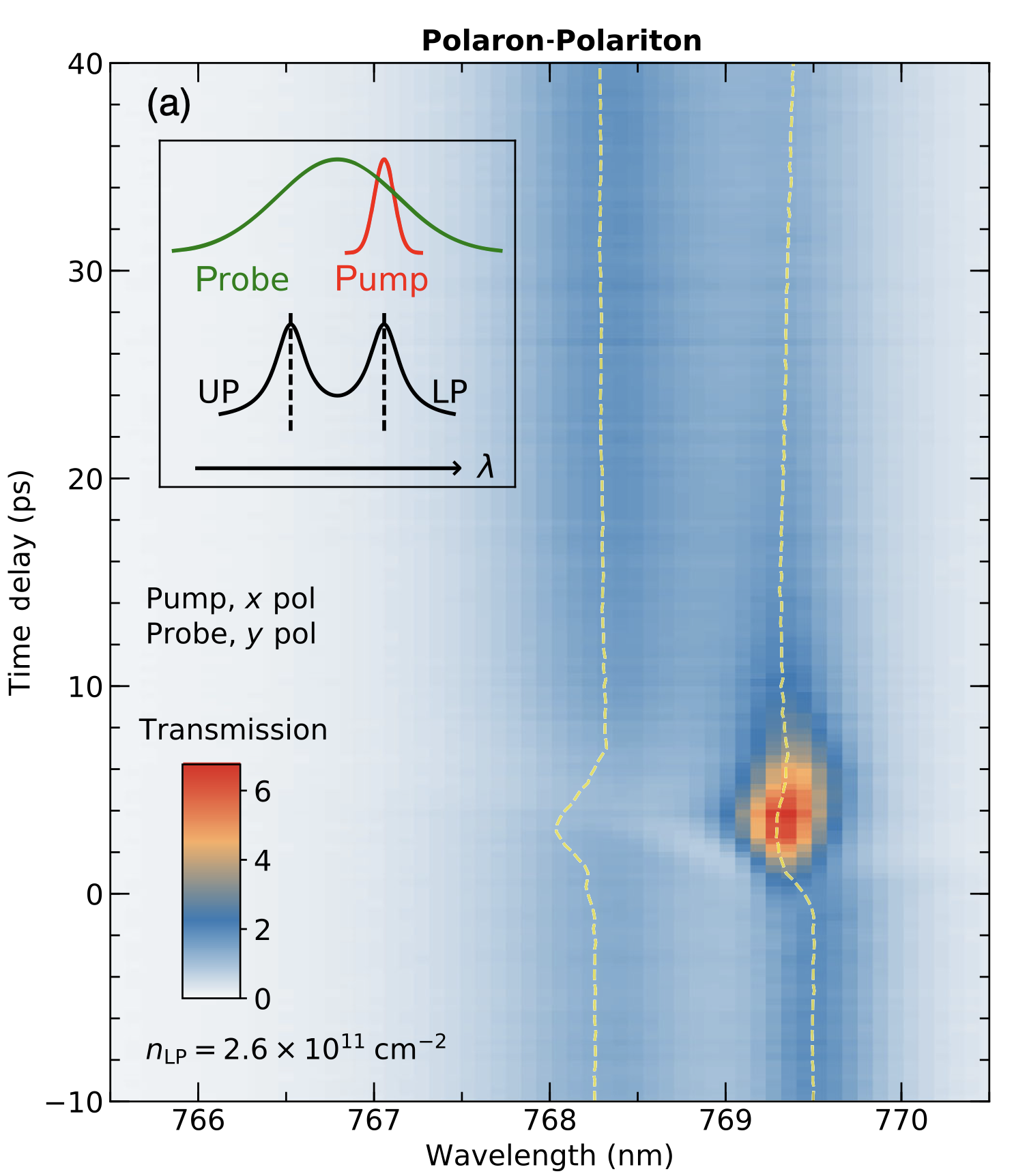}
\includegraphics[width=0.85\columnwidth]{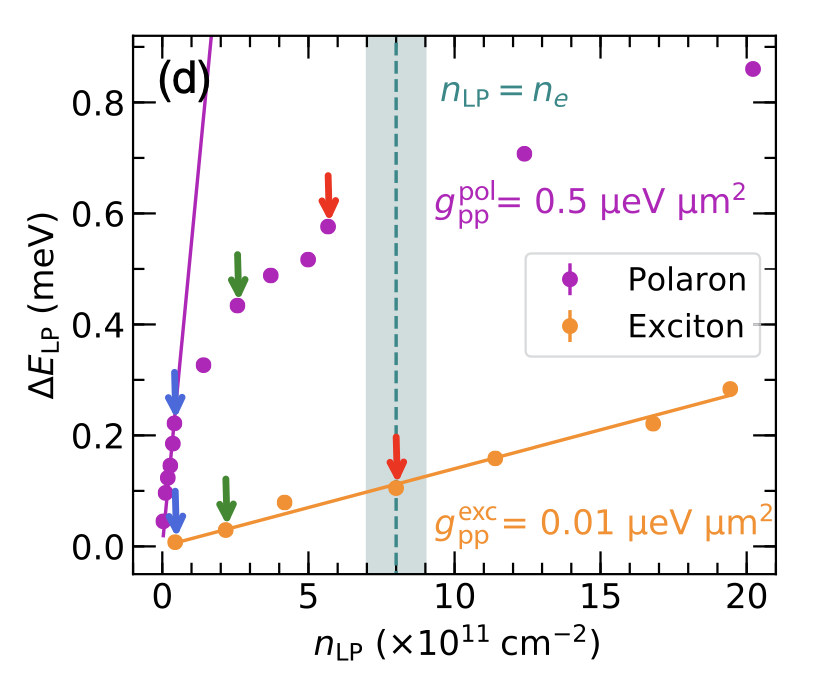}
\caption{Figure from Ref.~\cite{Tan2020}. (Top) Probe transmission spectrum for a typical pump-probe experiment. A narrow pump field injects resonantly polaron-polaritons whereas a delayed and broad probe field covers both polariton branches.  Yellow dashed lines illustrate the upper and lower polaron-polariton resonances as a function of time delay. (Bottom) Energy shift of the the exciton- and lower polaron-polariton as a function of the 
concentration of the lower polaron-polaritons. } 
\label{FigTan2020}
\end{figure}

 In a recent experiment  using  a semiconductor monolayer of MoSe$_2$,  %sandwiched between two boron nitride flakes,
 the presence of a 2DEG was found to have strong effects on the interaction between polaron-polaritons~\cite{Tan2020}. A 
 layer of graphene allowed for the tuning of the electron density and  the setup was embedded in a high-finesse optical cavity consisting of a distributed-Bragg-reflector to increase the photon-exciton coupling. 
 {First, a narrow pump beam tuned to the lower-polariton energy created polaritons, which, due to their excitonic component, interacted with electrons in the 2DEG. The presence of a trion state, i.e., a bound state between an exciton and an itinerant electron, led to an indirect coupling between the polaritons and the trion, forming strongly interacting polaron-polaritons~\cite{Sidler2016}. }
 A broad probe beam was then applied after a delay time to create and measure the energy of  
 polaron-polaritons in the presence of the polaron-polaritons already created by the pump beam. As seen in Fig.~\ref{FigTan2020} (top), a 
 transient blue shift of the measured polaron-polariton energy was observed for certain delay times between the pump and 
 probe pulses. 
 This blue shift was interpreted as  
a repulsive interaction between the   polariton-polaritons, which was observed to increase significantly  when a 2DEG was present 
as shown in Fig.~\ref{FigTan2020} (bottom).  Here the polaron-polariton interactions (purple dots) in the presence 
  of the 2DEG are $\sim 50$ times larger than the bare exciton-polariton interactions (orange dots). 
  Intriguingly, the observed repulsive nature of the polariton-polariton interaction is at odds with the results in 
  Sec.~\ref{Quasiparticles}, which predict  the interaction mediated by the electrons to be  attractive since the excitons to a good approximation can be regarded as point bosons. 
  
  To explain this result theoretically, a Chevy type ansatz~\cite{Chevy2006} for the many-body wave function, which has been shown to 
  explain the experimental results in the linear regime~\cite{Sidler2016},  
  was employed to study polaron-polaron interactions 
  \begin{gather}
  \label{ChevyPP}
    |\Psi_{\mathbf p}\rangle=\big(\psi^{c}_{\mathbf p}\hat c_{\mathbf p}^\dagger+\psi^{c}_{\mathbf p}\hat x^\dagger_{\mathbf p}+\sum_{\mathbf q,\mathbf k}\psi_{\mathbf p,\mathbf k,\mathbf p}\hat x^\dagger_{\mathbf p+\mathbf q-\mathbf k}\hat e^\dagger_{\mathbf k}\hat e_{\mathbf q}\big)|\text{FS}\rangle. 
  \end{gather}
Here, $\hat c_{\mathbf p}^\dagger/\hat x_{\mathbf p}^\dagger/\hat e_{\mathbf p}^\dagger$ creates a cavity photon/exciton/electron 
with momentum $\mathbf p$, and $\psi^{i}$ and $\psi_{\mathbf p,\mathbf k,\mathbf p}$ are variational parameters determined by minimising the energy. This ansatz creates an exciton-polariton and includes the 
exciton-electron interaction at the level of one particle-hole excitation  in the Fermi sea $|\text{FS}\rangle$.

Writing this  many-body wave function formally as  $|\Psi_{\mathbf p}\rangle=\hat a^\dagger_{\mathbf p}|\text{FS}\rangle$ where $\hat a_{\mathbf p}^\dagger$ creates a  polaron-polariton on top of the Fermi sea, the polaron-polariton interactions $U$ was estimated by 
  \begin{gather}
  \label{polariton_interactions}
    U=\frac{\langle \text{FS}|\hat a_{\mathbf 0}\hat a_{\mathbf 0}\hat H \hat a^\dagger_{\mathbf 0}\hat a^\dagger_{\mathbf 0}|\text{FS}\rangle}{\langle \text{FS}|\hat a_{\mathbf 0}\hat a_{\mathbf 0}\hat a^\dagger_{\mathbf 0}\hat a^\dagger_{\mathbf 0}|\text{FS}\rangle} -2E_0.
  \end{gather}
 Here,  $E_0$ is the energy of a single polaron-polariton and $\hat H$ the many-body Hamiltonian. Following this approach, a repulsive interaction was theoretically obtained and attributed to phase-space filling effects.

%  The repulsive interaction was attributed to phase-space filling effects and to the non-equilibrium character of the optical excitations.
Alternatively, a theoretical analysis using equilibrium Green's functions closely following the approach leading to the mediated interaction in 
Fig.~\ref{Fig1}(b), predicted an attractive interaction between the polaron-polaritons 
  mediated by the electrons, which is consistent with the results of Sec.~\ref{perturbation}~\cite{Bastarrachea2021,bastarrachea2021polaritons}. A blue shift of the polaron-polariton energy 
  could instead be explained by repulsive interactions with trions.

Strong non-linearities of polaritons immersed in an electron gas were observed in an experiment based on a MoSe$_2$ 
monolayer in an optical cavity~\cite{emmanuele2020highly}. The interactions between the electrons and the 
excitons were interpreted in terms of the formation of trion-polaritons instead of the formation of 
polaron-polaritons, although the coupling of trions to cavity photons is small. 
However, in the regime where the binding energy of the trion is much larger than the Fermi energy, these two interpretations 
yield almost the same predictions~\cite{Glazov2020}. 
Upon increasing the density of the excitons, two main effects appeared: A quench of the polaron-polariton oscillator
strength and positive and negative energy shifts of the lower and middle polaron-polaritons  $10-100$ times larger than for neutral exciton-polaritons. These observations were theoretically interpreted as a consequence of the underlying composite electron-hole nature of the excitons, which led to  phase-space filling effects. The exchange interaction between two trion-polaritons was predicted to be attractive and $\sim 5$ times stronger than 
the interaction between exciton-polaritons~\cite{song2022attractive}. In Ref.~\cite{Bastarrachea2021}, it was shown that equilibrium field theory
provides an alternative interpretation of the   signs as well as  magnitudes of the observed energy shifts in terms of interactions with trions.

Using multi-dimensional coherent spectroscopy on a monolayer WS$_2$~\cite{muir2022interactions}, interactions between 
excitons immersed in a 2DEG were studied at time scales shorter than the typical formation time of Fermi polarons.
 The observed interaction effects were attributed to phase-filling effects where excitons compete for the same electrons to form polarons.
 In addition, {in Ref.~\cite{muir2022interactions}}  a bipolaron involving excitons in different valleys with a rather large binding energy was found.

\subsection{Bose polaron-polariton interactions}
\begin{figure}
\includegraphics[width=0.99\columnwidth]{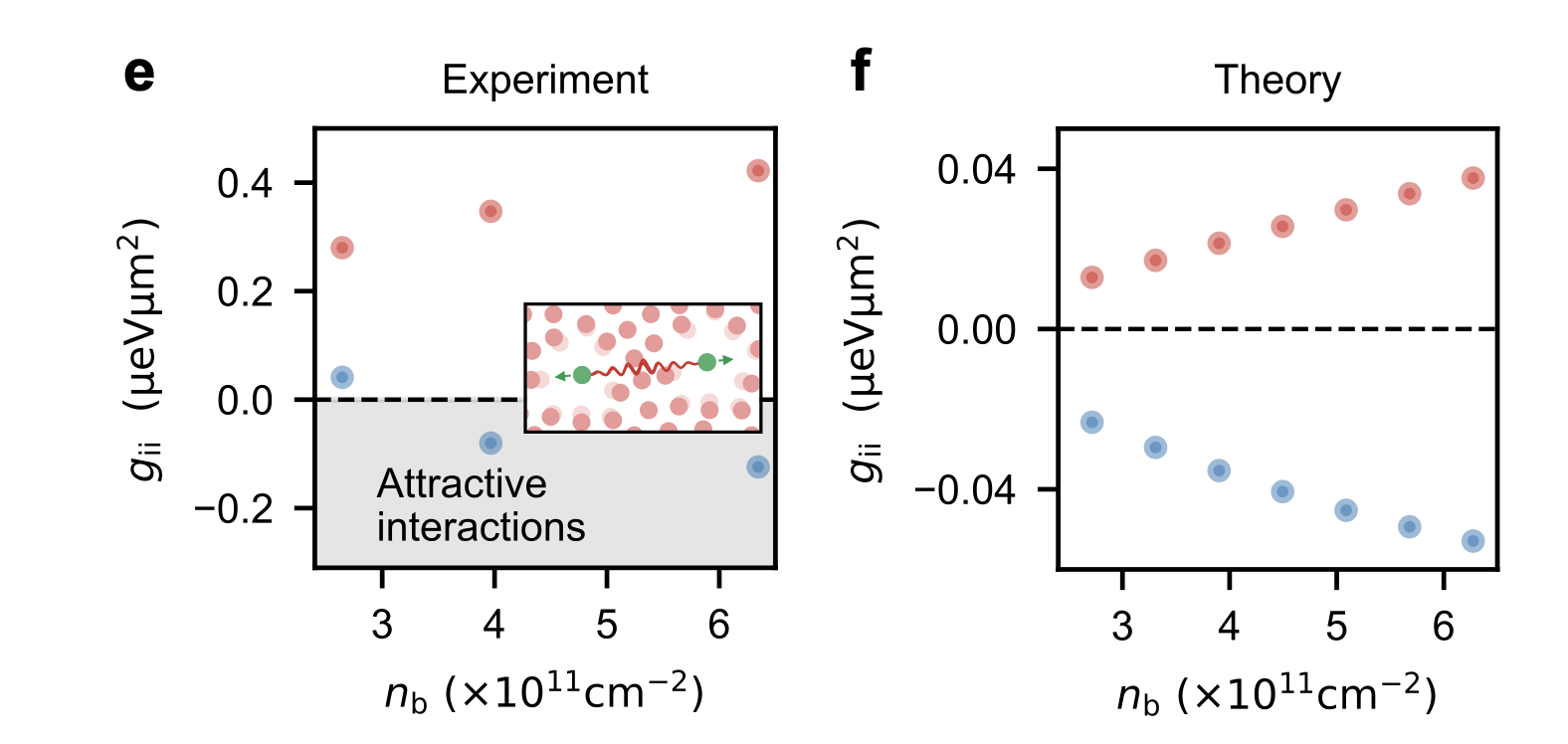}
\caption{Figure from Ref.~\cite{tan2022bose}. Left panel: Experimental interaction $g_{ii}$ between attractive (red dots) and repulsive (blue dots)
 polaron-polaritons as a function of the density $n_b$ of the majority excitons. Right panel: Theoretical 
results from a theory similar to Eq.~\eqref{polariton_interactions}.  }
\label{FigTan2022}
\end{figure}

Bose polaron-polaritons have also been created in 2D semiconductors using mixtures of exciton-polaritons in two spin 
states~\cite{Takemura2014,Takemura2017,Navadeh2019}. The presence of a bi-exciton state formed by two excitons with opposite spins  allowed for the demonstration of a polaritonic Feshbach resonance, leading to the formation of strongly interacting Bose polaron-polaritons. Theoretical predictions showed  good agreement with these initial experiments~\cite{Takemura2016,levinsen2019,Bastarrachea2019}. Recently, the interaction between such Bose polarons was
explored experimentally  in a monolayer MoSe$_2$ embedded in an optical cavity~\cite{tan2022bose}. 
Using the spin-valley optical selection rules to create a bosonic bath of spin-polarized polaritons  mixed with 
polaritons with opposite spin and a much smaller concentration,
 the energy of the resulting Bose polarons was measured as a function of their concentration. 
From this, a polaron-polaron interaction was extracted exhibiting a linear dependence on the density of the bath as 
shown in Fig.~\ref{FigTan2022}(left), which indicates that it is a mediated interaction. 
The mediated interaction was observed to be attractive for the attractive polaron-polariton and repulsive 
for the repulsive polaron-polariton. 
 A theoretical approach based on a variational ansatz similar to Eq.~\ref{ChevyPP} was shown to capture the sign-flip of the polariton interactions but predicted an interaction ten times smaller than the experimentally observed,  see Fig.~\ref{FigTan2022} (right)~\cite{tan2022bose}.
%%%%%%%%%%%%%%%%%%%%%%%%%%%%%%%%%%%%%%%%%%%%%%%%%%%%%%%%%%%%%%%%%%%%%%%%%%%
%%%%%%%%%%%%%%%%%%%%%%%%%%%%%%%%%%%%%%%%%%%%%%%%%%%%%%%%%%%%%%%%%%%%%%%%%%%

These experiments using  2D quantum materials have interesting results whose  relation to the interaction between 
quasiparticles in equilibrium remains to be clarified. 
 The understanding of these systems is  typically more complicated than their atomic gas counterparts
because the optical excitations can be far from equilibrium,  the microscopic interactions between excitons and electrons are not known precisely and not on a simple short range form, and the excitons are composite objects and thus not strictly bosonic particles.
This makes a quantitative understanding of the results including the links to Landau's quasiparticle interaction an open and  challenging
problem.

\section{Mediated collective phases} \label{manybodybose}
In this perspective, we have discussed the mediated interaction between individual quasiparticles. When the concentration 
of these quasiparticles is non-zero, such interactions can give rise to new phases and while this is 
an extensive topic of its own beyond the scope of this article, we here briefly outline some the recent results in the fields of atomic gases and 2D semiconductors.% focusing on superfluid/superconducting phases.

{\bf Boson-mediated $s$-wave superfluidity.-} Many studies have focused on Bose-Fermi mixtures and $s$-wave superfluidity via the exchange of phonon modes. The basic idea is that the bosons form a weakly interacting BEC, which mediates an attractive interaction between two distinguishable fermion 
species (spin $\uparrow,\downarrow$). Neglecting retardation effects and 
assuming weak short range interactions, one can derive analytical results using BCS-theory. In particular, 
the critical temperature is $k_BT_c=\frac{\gamma}{\pi}\left(\frac{2}{e}\right)^{7/3}\epsilon_Fe^{1/\lambda}$
with  $\lambda=\frac{m_ck_F}{2\pi^2\hbar^2}g_{\uparrow\downarrow}(1-g^2/g_{bb}g_{\uparrow\downarrow})$,
where $g_{\uparrow\downarrow}$ is the direct interaction between the $\uparrow$ and $\downarrow$ fermions, 
$g$ is the boson-fermion interaction, and $g_{bb}$ is the boson-boson interaction.~\cite{Viverit2002}. 
 This result stresses that the superfluid phase can be stabilised by increasing the fermion-boson coupling strength $g$ or by 
 softening the BEC with a smaller $g_{bb}$.
 
{\bf  Boson-mediated $p$-wave superfluidity.-} 
In a mixture of single spin fermions and bosons, there cannot be fermionic $s$-wave  pairing. The interaction mediated by the bosons can instead induce $p$-wave pairing, which was investigated in  early papers~\cite{Efremov2002,suzuki2008p}. This was extended to strong interactions using Eliashberg theory~\cite{Wang2006,kinnunen2018induced,matsyshyn2018p}, 
as well as to imbalanced Fermi mixtures~\cite{wang2019superfluid,wu2020effective,Dieckmann2008}. 
Topological $p$-wave pairing in a 2D gas of fermions mixed with a 3D weakly interacting BEC was analysed using strong coupling Eliashberg theory combined with Kosterlitz-Thouless theory~\cite{Wu2016}. In Refs.~\cite{Bulgac2006,bulgac2009induced,Patton2012},
$p$-wave pairing was  predicted to be favorable in spin-imbalanced Fermi mixtures where 
a fraction of the mixture forms a BEC of strongly bound dimers mediating an attractive interaction between  in the remaining 
spin polarised fermions.

{\bf Mediated superfluidity in optical lattices.-} 
There has  been extensive interest in exploring  
induced superfluidity of atoms in optical lattices. In Ref.~\cite{Illuminati2004}, the pairing of a 
two-component Fermi gas mixed with bosons  in a 3D optical lattice  was studied,
predicting an increased critical temperature compared to that coming from an on-site direct interaction.
In 2D optical lattices, the long-ranged character of the induced interaction mediated by weakly interacting
bosons can compete against a short-ranged  direct interaction between  fermions, which leads to a  rich phase
diagram, including $s$-, $p$-, and $d$-wave pairing together with charge and spin density ordering~\cite{Wang2005,Klironomos2007}.  Exotic superfluid phases in 2D optical lattices have been explored in the context of tunable 
boson-assisted finite-range interactions and  Majorana corner modes~\cite{Yu-Biao2023},  topological
superfluids~\cite{midtgaard2016topological,zhang2021stabilizing},  time-reversal-invariant topological
superfluids~\cite{midtgaard2017time}, pairing mediated by hard-core bosons~\cite{santiago2023collective}, FFLO
phases~\cite{singh2020enhanced}, $p$-wave pairing~\cite{wu2020effective}, and mixed dimensional mixtures~\cite{okamoto2017fermion} % \georg{This references also appears in previous section?}, and mixed-dimensional mixtures~\cite{okamoto2017fermion}.

{\bf Fermion-mediated collective phases.-}
The effects of spin-dependent fermion-mediated interactions on the properties of BECs such as stability, phase separation, and depletion were investigated in Refs.~\cite{Liao2020,Zheng2021}. In Refs.~\cite{Bo2020,Jiang2021}, the 
 intrinsic character of fermion-mediated interactions was explored, and exotic quantum many-body phases resulting from such interactions in optical lattices were proposed~\cite{Buchler2003,Arguello2022,de2014fermion}. The interaction  mediated by a superfluid two-component Fermi gas was shown to give rise to strong effects on the sound mode spectrum of a BEC, including 
avoided crossing features when the sound mode frequencies of the two superfluid match~\cite{Kinnunen2015}. In Ref.~\cite{Shen2024}, 
the effects of a fermion mediated interaction on the phase diagram of a Bose-Fermi mixture were  explored, 
and it was shown that strong coupling effects 
stabilise the mixture again collapse and phase separature well beyond what is expected from mean-field 
theory. These finding were in good agreement with recent experimental findings~\cite{Patel2023}.

\label{vanderWaalsmany}

{\bf Exciton-polariton mediated superconductivity.-}  The experimental realization of exciton-polariton BECs (or coherent states)
has motivated several investigations concerning how  they can mediate an attractive interaction between electrons
thereby inducing superconductivity in a solid state setting. 
Contrary to atomic gases where the interaction between the fermions can be suppressed, the mediated interaction between electrons has  to 
overcome their Coulomb repulsion. For low condensate densities, the net interaction is typically dominated by Coulomb repulsion, and electron pairing cannot take place. At high densities however, the attractive mediated interaction becomes strong enough to support Cooper pairs and superconductivity.
The first proposals studied a 2DEG sandwiched between  quantum wells in which a condensate of exctiton-polaritons can mediate an 
attractive interaction~\cite{Laussy2010,laussy2012superconductivity,Cotle2016}. 
With the arrival of TMDs, 
Bogoliubov mode mediated superconductivity  recieved renewed attention~\cite{Skopelitis2018,sun2021theory,Yang2021}. This includes 
the development of a strong-coupling theory of condensate-mediated superconductivity~\cite{Sun2021}, and the
demonstration of topological superconductivity with a high critical temperature due to the 
light  mass of the polaritons forming the condensate reducing retardation effect~\cite{Julku2022}.  Topological $p$-wave as well as $s$-wave superconductivity mediated by excitons interacting with electrons via a trion state 
was analysed in Refs.~\cite{zerba2023realizing,vonmilczewski2023superconductivity}.  
Bound states of two photons  have also been predicted to emerge as a consequence of an attractive interaction mediated by a polariton condensate~\cite{camacho2021mediated}. {Recently, spinor BECs have been proposed to enhance superconductivity in TMD materials and atomic systems~\cite{bighin2022}. }

In the opposite regime where the electron gas can be regarded as the majority compared to the excitons, it has been shown that they can induce a roton minimum in the dispersion of the exciton condensate~\cite{Shelykh2010}. The presence of a roton mode in ultracold quantum gases has led to the observation of  dipolar supersolids~\cite{chomaz2018observation,chomaz2019long,norcia2021two}, and similar effects have been suggested to exist for exciton condensates~\cite{Matuszewski2012}.
The link between fermion-mediated supersolidity and boson-mediated superconductivity has been studied highlighting the importance of the renormalized interaction between polaritons and electrons~\cite{Cotle2016,Plyashechnik2023}. A recent review on the topic of exciton-photon interactions in 2D microcavities highlights the interplay between 
many-body physics and photonics~\cite{Shang2023}.
Further details regarding atomic Bose-Fermi mixtures can be found in a recent excellent review~\cite{baroni2024quantum}.

\section{Conclusions and Outlook}\label{Conclusions}
In this perspective article, we discussed how mediated interactions are an intrinsic feature of quasiparticles and 
explained different theorerical approaches for calculating them. We then described how the advent highly flexible 
ultracold atomic gases, optical cavities,  and 2D TMDs has improved our theoretical as well as experimental understanding of such 
mediated interactions, and how they open ways to create  various interesting quantum phases. 

We end by discussing the many  interesting and open questions raised by these results. For instance, while the quasiparticle interaction 
between atomic Fermi polarons has been observed and is well understood for weak to moderately strong interactions, the experimental 
results for strong interactions are only partly explained by theory as discussed in Sec.~\ref{fermimediated}. 
Moreover, the quasiparticle interaction between atomic Bose 
polarons has not been observed so far, which  is somewhat surprising since a BEC is more compressible and therefore in general gives rise to a stronger mediated 
interaction as we saw in Sec.~\ref{quantumBosegases}. One possible reason is the rather short lifetime of the Bose polaron, 
which may wash out the effects of a quasiparticle interaction. The interplay between mediated interactions, few-body physics, and the 
formation of bound states is also an interesting topic that likely will unveil a new panorama to realise  exotic many-body phases. 
So far, studies of mediated interactions have furthermore mostly focused simple environments such as an ideal Fermi gas or a weakly 
interacting BEC, whereas  mediated interactions in strongly correlated phases  
remain relatively  unexplored. Understanding the puzzling results regarding interactions between excitons in TMDs mediated 
by electrons or other excitons is also a largely open and important problem. This is complicated by several factors such as 
the composite nature of the excitons leading to phase space filling effects, the non-trival nature of the exciton-electron and 
exciton-exciton interactions, and the inherent non-equilibrium nature of these systems, which may require the 
development of new theoretical frameworks. Understanding this can lead to the realisation of new quantum phases in a solid-state setting 
with far reaching perspectives. All this illustrates that  mediated interactions are a crucial concept in quantum many-body physics, and there remains a vast landscape of unexplored possibilities, which could pave the way for breakthroughs in science and technology.

{\it Acknowledgments.-} R. P acknowledges Grant No. IN117623 from DGAPA (UNAM) A. C. G. acknowledges financial support from Grant UNAM DGAPA PAPIIT No. IA101923, PAPIME No. PE101223 and PIIF 2023.  
G. B acknowledges the Danish National Research Foundation through the Center of Excellence CCQ (Grant no. DNRF156).

\bibliography{references}

\end{document}